\documentclass[journal]{IEEEtran}
\usepackage{mathrsfs}
\usepackage[centertags]{amsmath}
\usepackage{amsfonts}
\usepackage{amssymb}
\usepackage{amsthm}
\usepackage{cases}
\usepackage{indentfirst}
\usepackage{epsfig}
\usepackage{graphicx}
\usepackage{color}
\usepackage{CJK}
\usepackage{underscore}
\usepackage{url}
\usepackage[numbers,sort&compress]{natbib}
\usepackage{subfigure} 
\usepackage{caption}

\newtheorem{theorem}{Theorem}[section]
\newtheorem{corollary}{Corollary}[section]
\newtheorem{lemma}{Lemma}[section]

\newtheorem{remark}{Remark}[section]
\newtheorem{definition}{Definition}[section]
\numberwithin{equation}{section}

\allowdisplaybreaks[4]

\ifCLASSINFOpdf
\else
\fi
\begin{document}
%
\title{An Optimal Condition of Robust Low-rank Matrices Recovery}

\author{Jianwen Huang,~
        Jianjun Wang,~
        Feng Zhang,~
        and~Wendong Wang
\thanks{This work was supported by Natural Science Foundation of China (No. 61673015, 61273020), Fundamental Research Funds for the Central
Universities (No. XDJK2015A007, SWU1809002), Youth Science and technology talent development project (No. Qian jiao he KY zi [2018]313). \emph{(Corresponding author: Jianjun Wang.)}}
\thanks{J. Huang, F. Zhang and W. Wang are with School of Mathematics and Statistics, Southwest
University, Chongqing, 400715, China, and College of Artificial Intelligence, Southwest
University, Chongqing, 400715, China (e-mail: hjw1303987297@126.com;
zhangf@email.swu.edu.cn; d.sylan@foxmail.com).}
\thanks{J. Wang is with College of Artificial Intelligence, Southwest
University, Chongqing, 400715, China (e-mail: wjj@swu.edu.cn).}
}

\markboth{}%
{Shell \MakeLowercase{\textit{et al.}}: Bare Demo of IEEEtran.cls for IEEE Journals}

\maketitle
\begin{abstract}
In this paper we investigate the reconstruction conditions of nuclear norm minimization for low-rank matrix recovery. We obtain sufficient conditions $\delta_{tr}<t/(4-t)$ with $0<t<4/3$ to guarantee the robust reconstruction $(z\neq0)$ or exact reconstruction $(z=0)$ of all rank $r$ matrices $X\in\mathbb{R}^{m\times n}$ from $b=\mathcal{A}(X)+z$ via nuclear norm minimization. Furthermore, we not only show that when $t=1$, the upper bound of $\delta_r<1/3$ is the same as the result of Cai and Zhang \cite{Cai and Zhang}, but also demonstrate that the gained upper bounds concerning the recovery error are better. Moreover, we prove that the restricted isometry property condition is sharp. Besides, the numerical experiments are conducted to reveal the nuclear norm minimization method is stable and robust for the recovery of low-rank matrix.
\end{abstract}


\IEEEpeerreviewmaketitle

\section{Introduction}
\label{sec.1}

\IEEEPARstart{S}{uppose} that $X\in\mathbb{R}^{m\times n}$ is an unknown low rank matrix, $\mathcal{A}: \mathbb{R}^{m\times n}\to\mathbb{R}^q$ is a known linear map, $b\in\mathbb{R}^q$ is a given observation and $z\in\mathbb{R}^q$ is measurement error. The rank minimization problem is defined as follows:
\begin{align}\label{eq.1}
\min_{X}~\mbox{rank}(X)~\mbox{s.t.}~\|\mathcal{A}(X)-b\|_2\leq\epsilon,
\end{align}
where $b=\mathcal{A}(X)+z$ and $\epsilon$ stands for the noise level. Since the problem (\ref{eq.1}) is NP-hard in general, Recht et al. \cite{Recht et al} introduced a convex relaxation, which minimizes the nuclear norm (also known as the Schatten $1$-norm or trace norm)
\begin{align}\label{eq.2}
\min_{X}~\|X\|_*~\mbox{s.t.}~\|\mathcal{A}(X)-b\|_2\leq\epsilon,
\end{align}
where $\|X\|_*=\sum_i^{\min\{m,n\}}\sigma_i(X)$ and $\sigma_i(X)$ is the $i$-th largest  singular values of matrix $X$. The problem (\ref{eq.2}) is convex, thus there are a large number of approaches which can be used to solve it. Some researchers have developed fast algorithms for solving it, see \cite{Tanner and Wei,Zhang and Li,Lu et al,Wen et al 2017a,Guo et al,Wen et al 2017b,Chang et al 2018,Wen J et al 2019,Huang H et al 2018}.

When $m=n$ and the matrix $X=\mbox{diag}(x)~(x\in\mathbb{R}^m)$ is a diagonal matrix, the problems (\ref{eq.1}) and (\ref{eq.2}) degenerate to the $l_0$-minimization and $l_1$-minimization, respectively, which belong to the main optimization problems in compressed sensing (CS).

In order to study the relationship between the rank minimization problem and the nuclear norm minimization problem, Cand$\grave{e}$s and Plan \cite{Candes and Plan} extended the notion of restricted isometry constant proposed by Cand$\grave{e}$s \cite{Candes} to low-rank matrix recovery case. The concept is as follows:
\begin{definition}\label{def.1}
Let $\mathcal{A}: \mathbb{R}^{m\times n}\to\mathbb{R}^q$ be a linear map. For any integer $r~(1\leq r\leq\min\{m,n\})$, the restricted isometry constant (RIC) of order $r$ is defined as the smallest positive number $\delta_r$ that satisfies
\begin{align}\label{eq.3}
(1-\delta_r)\|X\|^2_F\leq\|\mathcal{A}(X)\|^2_2\leq(1+\delta_r)\|X\|^2_F
\end{align}
for all $r$-rank matrices $X$ (i.e., the rank of $X$ is at most $r$), where $\|X\|^2_F=\left<X,X\right>=Tr(X^{\top}X)$ is the Frobenius norm of $X$, which is also equal to the sum of the square of singular values and the inner product in $\mathbb{R}^{m\times n}$ as $\left<X,Y\right>=Tr(X^{\top}Y)=\sum^m_{i=1}\sum^n_{j=1}X_{ij}Y_{ij}$ for matrices $X$ and $Y$ of the same dimension.
\end{definition}
By the aforementioned definition, it is easy to see that if $r_1\leq r_2$, then $\delta_{r1}\leq\delta_{r2}$.

Although it is not easy to examine the restricted isometry property for a given linear map, it is one of the central notions in low-rank matrix recovery. In fact, it has been showed \cite{Recht et al} that Gaussian or sub-Gaussian random measurement map $\mathcal{A}$ fulfills the restricted isometry property with high probability.

There exist many sufficient conditions based on RIP for the exact recovery (i.e., in the case of $z=0$) of any rank-$r$ matrices through the formulation (\ref{eq.2}). These contain $\delta_{5r}<1/10$ \cite{Recht et al}, $\delta_{4r}<\sqrt{2}-1$ \cite{Candes and Plan}, $\delta_{4r}<0.558$ \cite{Mohan and Fazel}, $\delta_{r}<1/3$ \cite{Cai and Zhang}, and $\delta_{2r}<\sqrt{2}/2$ \cite{Wang and Li}. For other related works, see, e.g., \cite{Kong and Xiu,Chen and Li,Wen J M et al,Chang et al 2017,Wen J M et al b,Wang Y et al,Fang et al,Wang et al,Ge et al}.

In special, Cai and Zhang \cite{Cai and Zhang 2014} showed that for any given $t\geq4/3$, $\delta_{tr}<\sqrt{(t-1)/t}$ ensures the exact reconstruction for all matrices with rank no more than $r$ in the noise-free case via the constrained nuclear norm minimization (\ref{eq.2}). Furthermore, for any $\varepsilon>0$, $\delta_{tr}<\sqrt{(t-1)/t}+\varepsilon$ doesn't suffice to make sure the exact recovery of all $r$-rank matrices for large $r$. Besides, they showed that condition
$\delta_{tr}<\sqrt{(t-1)/t}$ suffices for robust reconstruction of nearly low-rank matrices in the noisy case.

Motivated by the aforementioned papers, we further discuss the upper bounds of $\delta_{tr}$ associated with some linear map $\mathcal{A}$ as $0<t<4/3$. Sufficient conditions regarding $\delta_{tr}$ with $0<t<4/3$ are established to guarantee the robust reconstruction $(\epsilon\neq0)$ or $(\epsilon=0)$ of all $r$-rank matrices $X\in \mathbb{R}^{m\times n}$ satisfying $b=\mathcal{A}(X)+z$ with $\|z\|_2\leq\epsilon$ and $\|\mathcal{A}^*(z)\|\leq\epsilon$, respectively. Thereby, combined with \cite{Cai and Zhang 2014}, a complete description for sharp restricted isometry property (RIP) constants for all $t>0$ is established to ensure the exact reconstruction of all matrices with rank no more than $r$ via nuclear norm minimization.

The construction of this paper is as follows. In Section \ref{sec.2}, we will provide some fundamental lemmas that be employed. We present the main results and the proofs in Sections \ref{sec.3} and \ref{sec.5}, respectively. A series of numerical simulation experiments on low-rank matrix reconstruction are carried out in Section \ref{sec.4}. Lastly, conclusion is drawn in in Section \ref{sec.6}.

\section{Preliminaries}
\label{sec.2}

We begin by introducing basic notations. We also gather a few lemmas needed for the proofs of main results.

For any matrix $X\in\mathbb{R}^{m\times n}$, we assume w.o.l.g. that $m\leq n$, and
 the singular value decomposition (SVD) of $X$ is represented by
$$X=U\mbox{diag}(\sigma(X))V^{\top},$$
where $U\in\mathbb{R}^{m\times m}$ and $V\in\mathbb{R}^{n\times n}$ are orthogonal matrices, and $\sigma(X)=(\sigma_1(X),\cdots,\sigma_m(X))^{\top}$ indicates the vector of the singular values of $X$. Assume that $\sigma_1(X)\geq\sigma_2(X)\geq\cdots\geq\sigma_m(X)$. Consequently, the best $r$-rank approximation to the matrix $X$ is

$$
X^{(r)}=U\left[\begin{matrix}
\mbox{diag}(\sigma^r(X))&0\\
0&0
\end{matrix}\right]V^{\top},
$$
where $\sigma^r(X)=(\sigma_1(X),\cdots,\sigma_r(X))^{\top}$.

For a linear map $\mathcal{A}:~\mathbb{R}^{m\times n}\to\mathbb{R}^q$, denote by its adjoint operator $\mathcal{A}^*:~\mathbb{R}^q\to\mathbb{R}^{m\times n}$. Then, for all $X\in\mathbb{R}^{m\times n}$ and $b\in\mathbb{R}^q$, $\left<X,\mathcal{A}^*(b)\right>=\left<\mathcal{A}(X),b\right>$.

Without loss of generality, let $X$ be the original matrix that we want to find and $X^*$ be an optimal solution to the problem (\ref{eq.2}). Let $Z=X-X^*$. Let SVD of $U^{\top}ZV\in\mathbb{R}^{m\times m}$ be provided by
$$
U^{\top}ZV=U_0\left[\begin{matrix}
\mbox{diag}\left(\sigma_T(U^{\top}ZV)\right)&0\\
0&\mbox{diag}\left(\sigma_{T^c}(U^{\top}ZV)\right)
\end{matrix}\right]V^{\top}_0
$$
where $U_0,~V_0\in\mathbb{R}^{m\times m}$ are orthogonal matrices, $\sigma_T(U^{\top}ZV)=\left(\sigma_1(U^{\top}ZV),\cdots,\sigma_r(U^{\top}ZV)\right)^{\top}$, $\sigma_{T^c}(U^{\top}ZV)=\left(\sigma_{r+1}(U^{\top}ZV),\cdots,\sigma_m(U^{\top}ZV)\right)^{\top}$, and we suppose that $\sigma_1(U^{\top}ZV)\geq\cdots\geq\sigma_r(U^{\top}ZV)\geq\sigma_{r+1}(U^{\top}ZV)
\geq\cdots\geq\sigma_m(U^{\top}ZV)$. Therefore, the matrix $Z$ is decomposed as
$$Z=Z^{(r)}+Z^{(r)}_c,$$
where
$$
Z^{(r)}=UU_0\left[\begin{matrix}
\mbox{diag}\left(\sigma_T(U^{\top}ZV)\right)&0\\
0&0
\end{matrix}\right]V^{\top}_0V^{\top}
$$
and
$$
Z^{(r)}_c=UU_0\left[\begin{matrix}
0&0\\
0&\mbox{diag}\left(\sigma_{T^c}(U^{\top}ZV)\right)
\end{matrix}\right]V^{\top}_0V^{\top}.
$$
It is not hard to see that $X^{(r)}(Z^{(r)}_c)^{\top}=0$ and $(X^{(r)})^{\top}Z^{(r)}_c=0$.

In order to show the main results, we need some elementary identities, which were given in \cite{Zhang and Li 2018} (see Lemma $1$).
\begin{lemma}\label{le.1}
Give matrices $\{V_i:~i\in T\}$ in a matrix space $\mathcal{V}$ with inner product $\left<\cdot\right>$, where $T$ denotes the index set with $|T|=r$. Select all subsets $T_i\subset T$ with $|T_i|=k$, $i\in I$ and $|I|=(^r_k)$, then we get
\begin{align}\label{eq.4}
\sum_{i\in I}\sum_{p\in T_i}V_p=\left(\begin{matrix}
r-1\\
k-1
\end{matrix}\right)\sum_{p\in T}V_p~(k\geq1),
\end{align}
and
\begin{align}\label{eq.5}
\sum_{i\in I}\sum_{p\neq q\in T_i}\left<V_p,V_q\right>=\left(\begin{matrix}
r-2\\
k-2
\end{matrix}\right)\sum_{p\neq q\in T}\left<V_p,V_q\right>~(k\geq2).
\end{align}
\end{lemma}
Cai and Zhang developed a new elementary technique which states an elementary geometric fact: Any point in a polytope can be represented as a convex combination of sparse vectors (see Lemma $1.1$ in \cite{Cai and Zhang 2014}). It gives a crucial technical tool for the proof of our main results. It is also the special case $p=1$ of Zhang and Li's result (see Lemma $2.2$ in \cite{Zhang and Li 2018b}).
\begin{lemma}\label{le.2}
Let $r\leq m$ be an integer, and $\alpha$ be a positive real number. We can represent any vector $x$ in the set
$$V=\{x\in\mathbb{R}^m:~\|x\|_1\leq r\alpha,~\|x\|_{\infty}\leq\alpha\},$$
as a convex combination of $r$-sparse vectors, i.e.,
$$x=\sum_i\lambda_iu_i$$
where $\sum_i\lambda_i=1$ with $\lambda_i\geq0$, $|\sup(u_i)|\leq r$, $\sup(u_i)\subset\sup(x)$ and $\sum_i\lambda_i\|u_i\|^2_2\leq r\alpha^2.$
\end{lemma}
\begin{lemma}\label{le.3}
(Lemma $2.3$ in \cite{Recht et al}) Let $X,~Y$ be the matrices of same dimensions. If $XY^{\top}=0$ and $X^{\top}Y=0$, then
\begin{align}\label{eq.6}
\|X+Y\|_*=\|X\|_*+\|Y\|_*.
\end{align}
\end{lemma}

\begin{lemma}\label{le.4}
We have
\begin{align}\label{eq.7}
\|Z^{(r)}_c\|_*\leq\|Z^{(r)}\|_*+2\|X-X^{(r)}\|_*.
\end{align}
\end{lemma}
\noindent \textbf{Proof.}
Since $X^*$ is the optimal solution to the problem (\ref{eq.2}), we get
\begin{align}\label{eq.8}
\|X\|_*\geq\|X^*\|_*=\|X-Z\|_*.
\end{align}
Applying the reverse inequality to (\ref{eq.8}), we get
\begin{align}\label{eq.9}
\notag\|X-Z\|_*&=\|(X^{(r)}-Z^{(r)}_c)+(X-X^{(r)}-Z^{(r)})\|_*\\
&\geq\|X^{(r)}-Z^{(r)}_c\|_*-\|X-X^{(r)}-Z^{(r)}\|_*.
\end{align}
By Lemma $\ref{le.3}$ and the forward inequality, we get
\begin{align}\label{eq.10}
\notag&\|X^{(r)}+(-Z^{(r)}_c)\|_*-\|X-X^{(r)}+(-Z^{(r)})\|_*\\
&\geq\|X^{(r)}\|_*+\|Z^{(r)}_c\|_*-\|X-X^{(r)}\|_*-\|Z^{(r)}\|_*.
\end{align}
Combining with (\ref{eq.8}), (\ref{eq.9}) and (\ref{eq.10}), we get
\begin{align}
\notag\|Z^{(r)}_c\|_*&\leq\|X\|_*-\|X^{(r)}\|_*+\|X-X^{(r)}\|_*+\|Z^{(r)}\|_*\\
\notag&\leq\|Z^{(r)}\|_*+2\|X-X^{(r)}\|_*.
\end{align}
The proof of the lemma is completed.

\qed

Select positive integers $a$ and $b$ satisfying $a+b=tr$ and $b\leq a\leq r$. We use $T_i,~S_j$ to represent all possible index set contained in $\{1,2,\cdots,r\}$ (i.e., $T_i,~S_j\subset \{1,\cdots,r\}$) and $|T_i|=a$, $|S_j|=b$, where $i\in A$ and $j\in B$ with $|A|=(^r_a)$ and $|B|=(^r_b)$. Define
$$
Z^{(r)}_{T_i}=UU_0\left[\begin{matrix}
\mbox{diag}\left(\sigma_{T_i}(U^{\top}ZV)\right)&0\\
0&0
\end{matrix}\right]V^{\top}_0V^{\top},
$$
and
$$
Z^{(r)}_{S_j}=UU_0\left[\begin{matrix}
\mbox{diag}\left(\sigma_{S_j}(U^{\top}ZV)\right)&0\\
0&0
\end{matrix}\right]V^{\top}_0V^{\top}.
$$
Here $\sigma_{T_i}(U^{\top}ZV)$ $(\sigma_{S_j}(U^{\top}ZV))$ denotes the vector that equals to $\sigma_{T}(U^{\top}ZV)$ on $T_i$ $(S_j)$, and zero elsewhere.

\begin{lemma}\label{le.5}
We have
$$Z^{(r)}_c=\sum_k\mu_kU_k,~Z^{(r)}_c=\sum_k\nu_kV_k,~Z^{(r)}_c=\sum_k\tau_kW_k,$$
where $\sum_k\mu_k=\sum_k\nu_k=\sum_k\tau_k=1$ with $\nu_k,~\mu_k,~\tau_k\geq0$, $U_k,~V_k,~W_k$ are $b$-rank, $a$-rank and $(t-1)r$-rank $(t>1)$ with
\begin{align}\label{eq.13}
\sum_k\mu_k\|U_k\|^2_F\leq\frac{r^2}{b}\alpha^2,
\end{align}
\begin{align}\label{eq.14}
\sum_k\nu_k\|V_k\|^2_F\leq\frac{r^2}{a}\alpha^2,
\end{align}
and
\begin{align}\label{eq.15}
\sum_k\tau_k\|W_k\|^2_F\leq\frac{r^2}{t-1}\alpha^2.
\end{align}

\end{lemma}
\noindent \textbf{Proof.}
Set $$\alpha=\frac{\|Z^{(r)}\|_*+2\|X-X^{(r)}\|_*}{r}.$$
By Lemma $\ref{le.4}$, then
\begin{align}
\notag\|Z^{(r)}_c\|_*\leq r\alpha.
\end{align}
By the definition of $Z^{(r)}_c$, we get
\begin{align}\label{eq.11}
\|\sigma_{T^c}(U^{\top}ZV)\|_1\leq r\alpha\leq b\frac{r}{b}\alpha.
\end{align}
By the decomposition of $Z$, we get
\begin{align}\label{eq.12}
\notag\|\sigma_{T^c}(U^{\top}ZV)\|_{\infty}&\leq\frac{\|\sigma_{T}(U^{\top}ZV)\|_1}{r}\\
\notag&\leq\frac{\|Z^{(r)}\|_*+2\|X-X^{(r)}\|_*}{r}\\
&\leq\alpha\leq\frac{r}{b}\alpha.
\end{align}
Combining with Lemma $\ref{le.2}$, (\ref{eq.11}) and (\ref{eq.12}), $\sigma_{T^c}(U^{\top}ZV)$ is decomposed into the convex combination of $b$-sparse vectors, i.e., $\sigma_{T^c}(U^{\top}ZV)=\sum_k\mu_ku_k$ with
\begin{align}
\notag\sum_k\mu_k\|u_k\|^2_2\leq\frac{r^2}{b}\alpha^2.
\end{align}
Define
$$
U_k=UU_0\left[\begin{matrix}
0&0\\
0&\mbox{diag}(u_k)
\end{matrix}\right]V^{\top}_0V^{\top}.
$$
It is easy to see that $U_k$ is $b$-rank. Therefore, $Z^{(r)}_c$ is decomposed as
$Z^{(r)}_c=\sum_k\mu_kU_k$ with
\begin{align}
\notag\sum_k\mu_k\|U_k\|^2_F=\sum_k\mu_k\|u_k\|^2_2\leq\frac{r^2}{b}\alpha^2.
\end{align}
Likewise, $Z^{(r)}_c$ can also be denoted by
$$Z^{(r)}_c=\sum_k\nu_kV_k,~Z^{(r)}_c=\sum_k\tau_kW_k,$$
where $V_k$ is $a$-rank, $W_k$ is $(t-1)r$-rank $(t>1)$ with
\begin{align}
\notag\sum_k\nu_k\|V_k\|^2_F\leq\frac{r^2}{a}\alpha^2,
\end{align}
and
\begin{align}
\notag\sum_k\tau_k\|V_k\|^2_F\leq\frac{r^2}{t-1}\alpha^2.
\end{align}

\qed

One can easily check that $\left<Z^{(r)}_{T_i},U_k\right>=0$, $\left<Z^{(r)}_{S_j},V_k\right>=0$ and $\left<Z^{(r)},W_k\right>=0$.

\begin{lemma}\label{le.6}
We have that
for $0<t<1$,
\begin{align}\label{eq.16}
\notag&\frac{\rho_{a,b}(t)}{(^r_a)(^{r-a}_b)}\sum_{T_i\bigcap S_j=\emptyset}\bigg[\left\|\mathcal{A}\left(Z^{(r)}_{T_i}+Z^{(r)}_{S_j}\right)\right\|^2_2\\
\notag&-\frac{r-a-b}{abr}\left\|\mathcal{A}\left(bZ^{(r)}_{T_i}-aZ^{(r)}_{S_j}\right)\right\|^2_2\bigg]\\
&=-2t^2(2-t)ab\left<\mathcal{A}Z^{(r)},\mathcal{A}Z\right>+t\Delta_{a,b},
\end{align}
and for $1\leq t<4/3$,
\begin{align}\label{eq.17}
\notag&\rho_{a,b}(t)\sum_k\tau_k\bigg[\left\|\mathcal{A}\left(Z^{(r)}+(t-1)W_k\right)\right\|^2_2\\
\notag&-\left\|(t-1)\mathcal{A}\left(Z^{(r)}-W_k\right)\right\|^2_2\bigg]\\
&=-2t^3[ab-(t-1)r^2]\left<\mathcal{A}Z^{(r)},\mathcal{A}Z\right>+(4-3t)\Delta_{a,b},
\end{align}
where
$$\rho_{a,b}(t)=(a+b)^2-2ab(4-t),$$
and
\begin{align}\label{eq.18}
\notag\Delta_{a,b}=&\frac{r-b}{a(^r_a)}\sum_{i\in A,~k}\mu_k\bigg[a^2\left\|\mathcal{A}\left(Z^{(r)}_{T_i}+\frac{b}{r}U_k\right)\right\|^2_2\\
\notag&-b^2\left\|\mathcal{A}\left(Z^{(r)}_{T_i}-\frac{a}{r}U_k\right)\right\|^2_2\bigg]\\
\notag&+\frac{r-a}{b(^r_b)}\sum_{j\in B,~k}\nu_k\bigg[b^2\left\|\mathcal{A}\left(Z^{(r)}_{S_j}+\frac{a}{r}V_k\right)\right\|^2_2\\
&-a^2\left\|\mathcal{A}\left(Z^{(r)}_{S_j}-\frac{b}{r}V_k\right)\right\|^2_2\bigg].
\end{align}
\end{lemma}

\noindent \textbf{Proof.}
The proof takes advantage of the ideas from \cite{Cai and Zhang}, \cite{Zhang and Li 2018}. By Lemma $\ref{le.1}$, we get
\begin{align}\label{eq.45}
\notag\Delta_{a,b}&=(a^2-b^2)\bigg[\frac{r-b}{a(^r_a)}\sum_{i\in A}\|\mathcal{A}Z^{(r)}_{T_i}\|^2_2-\frac{r-a}{b(^r_b)}\sum_{j\in B}\|\mathcal{A}Z^{(r)}_{S_j}\|^2_2\bigg]\\
\notag&+\frac{2(a^2b+ab^2)}{r}\times\\
\notag&\left<\frac{r-b}{a(^r_a)}\sum_{i\in A}\mathcal{A}Z^{(r)}_{T_i}+\frac{r-a}{b(^r_b)}\sum_{j\in B}\mathcal{A}Z^{(r)}_{S_j},\mathcal{A}Z^{(r)}_c\right>\\
\notag=&(a^2-b^2)\bigg(\frac{r-b}{a(^r_a)}(^{r-1}_{a-1})\|\mathcal{A}Z^{(r)}\|^2_2\\
\notag&-\frac{r-a}{b(^r_b)}(^{r-1}_{b-1})\|\mathcal{A}Z^{(r)}\|^2_2\bigg)+\frac{2ab(a+b)}{r}\times\\
\notag&\left<\frac{r-b}{a(^r_a)}(^{r-1}_{a-1})\mathcal{A}Z^{(r)}
+\frac{r-a}{b(^r_b)}(^{r-1}_{b-1})\mathcal{A}Z^{(r)},\mathcal{A}Z^{(r)}_c\right>\\
\notag=&(a^2-b^2)\frac{a-b}{r}\|\mathcal{A}Z^{(r)}\|^2_2\\
\notag&+2abt\frac{2r-a-b}{r}\left<\mathcal{A}Z^{(r)},\mathcal{A}Z^{(r)}_c\right>\\
=&t\rho_{a,b}(t)\|\mathcal{A}Z^{(r)}\|^2_2+2abt(2-t)\left<\mathcal{A}Z^{(r)},\mathcal{A}Z\right>.
\end{align}
where the first equality follows from Lemma $\ref{le.5}$, i.e., $Z^{(r)}_c$ has the convex decomposition, and in the second equality, we used the identity (\ref{eq.4}).

As $0<t<1$, by Lemma $2$ in \cite{Zhang and Li 2018}, we get
\begin{align}\label{eq.46}
\notag LHS&=\rho_{a,b}(t)\left(\frac{a+b}{r}\right)^2\|\mathcal{A}Z^{(r)}\|^2_2\\
&=\rho_{a,b}(t)t^2\|\mathcal{A}Z^{(r)}\|^2_2.
\end{align}
Substituting (\ref{eq.45}) to the right hand side of (\ref{eq.16}), we get
\begin{align}
\notag RHS=&t\bigg[t\rho_{a,b}(t)\|\mathcal{A}Z^{(r)}\|^2_2+2abt(2-t)\left<\mathcal{A}Z^{(r)},\mathcal{A}Z\right>\bigg]\\
\notag&-2t^2(2-t)ab\left<\mathcal{A}Z^{(r)},\mathcal{A}Z\right>\\
\notag=&LHS.
\end{align}
Accordingly, the identity (\ref{eq.16}) holds.

As $1\leq t<4/3$, we get
\begin{align}
\notag LHS=&\rho_{a,b}(t)\bigg\{[1-(t-1)^2]\|\mathcal{A}Z^{(r)}\|^2_2\\
\notag&+2(t-1)t\left<\mathcal{A}Z^{(r)},\sum_k\tau_k\mathcal{A}W_k\right>\bigg\}\\
\notag=&\rho_{a,b}(t)\bigg\{[1-(t-1)^2]\|\mathcal{A}Z^{(r)}\|^2_2\\
\notag&+2(t-1)t\left<\mathcal{A}Z^{(r)},\mathcal{A}Z^{(r)}_c\right>\bigg\}\\
\notag=&\rho_{a,b}(t)\bigg\{(4t-3t^2)\|\mathcal{A}Z^{(r)}\|^2_2\\
\notag&+2(t-1)t\left<\mathcal{A}Z^{(r)},\mathcal{A}Z\right>\bigg\}.
\end{align}
We have
\begin{align}
\notag RHS=&(4-3t)\bigg[t\rho_{a,b}(t)\|\mathcal{A}Z^{(r)}\|^2_2\\
\notag&+2abt(2-t)\left<\mathcal{A}Z^{(r)},\mathcal{A}Z\right>\bigg]\\
\notag&-2t^3[ab-(t-1)r^2]\left<\mathcal{A}Z^{(r)},\mathcal{A}Z\right>\\
\notag=&(4t-3t^2)\rho_{a,b}(t)\|\mathcal{A}Z^{(r)}\|^2_2
+2t\bigg\{ab(2-t)(4-3t)\\
\notag&-t^2[ab-(t-1)r^2]\bigg\}\left<\mathcal{A}Z^{(r)},\mathcal{A}Z\right>\\
\notag=&LHS.
\end{align}
Therefore, the identity (\ref{eq.17}) holds.

\qed

\begin{lemma}\label{le.7}
It holds that
\begin{align}\label{eq.19}
\notag&\{[(a+b)^2-4ab]t-[(a+b)^2-2ab](2-t)\delta_{tr}\}\|Z^{(r)}\|^2_F\\
&-2abr\delta_{tr}\alpha^2(2-t)\leq\Delta_{a,b}.
\end{align}
\end{lemma}

\noindent \textbf{Proof.}
Note that the ranks of matrices $U_k,~Z^{(r)}_{S_j}$ are no more than $b$, the ranks of matrices $V_k,~Z^{(r)}_{T_i}$ are at most $a$ and $a+b=tr$. By the $tr$-order restricted isometry property, we get
\begin{align}
\notag\Delta_{a,b}\geq&\frac{r-b}{a(^r_a)}\sum_{i\in A,~k}\mu_k\bigg[a^2(1-\delta_{tr})\left\|Z^{(r)}_{T_i}+\frac{b}{r}U_k\right\|^2_F\\
\notag&-b^2(1+\delta_{tr})\left\|Z^{(r)}_{T_i}-\frac{a}{r}U_k\right\|^2_F\bigg]\\
\notag&+\frac{r-a}{b(^r_b)}\sum_{j\in B,~k}\nu_k\bigg[b^2(1-\delta_{tr})\left\|Z^{(r)}_{S_j}+\frac{b}{r}V_k\right\|^2_F\\
\notag&-a^2(1+\delta_{tr})\left\|Z^{(r)}_{S_j}-\frac{b}{r}V_k\right\|^2_F\bigg].
\end{align}
Since the inner product of $Z^{(r)}_{T_i}~(Z^{(r)}_{S_j})$ and $U_k~(V_k)$ equals to zero, by some elementary calculation, we get
\begin{align}\label{eq.20}
\notag\Delta_{a,b}\geq&(a^2-b^2)\left[\frac{r-b}{a(^r_a)}\sum_{i\in A}\|Z^{(r)}_{T_i}\|^2_F-\frac{r-a}{b(^r_b)}\sum_{j\in B}\|Z^{(r)}_{S_j}\|^2_F\right]\\
\notag&-(a^2+b^2)\delta_{tr}\bigg[\frac{r-b}{a(^r_a)}\sum_{i\in A}\|Z^{(r)}_{T_i}\|^2_F\\
\notag&+\frac{r-a}{b(^r_b)}\sum_{j\in B}\|Z^{(r)}_{S_j}\|^2_F\bigg]-\frac{2ab^2(r-b)\delta_{tr}}{r^2}\sum_k\mu_k\|U_k\|^2_F\\
&-\frac{2a^2b(r-a)\delta_{tr}}{r^2}\sum_k\nu_k\|V_k\|^2_F.
\end{align}
By Lemma $\ref{le.1}$, we get
\begin{align}\label{eq.21}
\sum_{i\in A}\|Z^{(r)}_{T_i}\|^2_F=(^{r-1}_{a-1})\|Z^{(r)}\|^2_F,
\end{align}
and
\begin{align}\label{eq.22}
\sum_{j\in B}\|Z^{(r)}_{S_j}\|^2_F=(^{r-1}_{b-1})\|Z^{(r)}\|^2_F.
\end{align}
Substituting (\ref{eq.21}) and (\ref{eq.22}) into (\ref{eq.20}) and combining with inequalities (\ref{eq.13}) and (\ref{eq.14}), we get
\begin{align}
\notag\Delta_{a,b}\geq&\frac{(a-b)^2(a+b)}{r}\|Z^{(r)}\|^2_F-(a^2+b^2)\delta_{tr}(2-t)\|Z^{(r)}\|^2_F\\
\notag&-\frac{2ab^2(r-b)\delta_{tr}}{r^2}\frac{r^2\alpha^2}{b}
-\frac{2a^2b(r-a)\delta_{tr}}{r^2}\frac{r^2\alpha^2}{a}\\
\notag=&\{[(a+b)^2-4ab]t-[(a+b)^2-2ab](2-t)\delta_{tr}\}\|Z^{(r)}\|^2_F\\
\notag&-2abr\delta_{tr}\alpha^2(2-t).
\end{align}

\qed

\begin{lemma}\label{le.9}
(Lemma $4.1$ in \cite{Cai and Zhang}) For all linear maps $\mathcal{A}:~\mathbb{R}^{m\times n}\to\mathbb{R}^q$ and $r\geq2$, $s\geq2$, we have
\begin{align}\label{eq.24}
\delta_{sr}\leq(2s-1)\delta_r.
\end{align}
\end{lemma}

\begin{lemma}\label{le.10}
It holds that
for $0<t<1$,
\begin{align}\label{eq.25}
\notag&\frac{\rho_{a,b}(t)}{(^r_a)(^{r-a}_b)}\sum_{T_i\bigcap S_j=\emptyset}\bigg[\left\|\mathcal{A}\left(Z^{(r)}_{T_i}+Z^{(r)}_{S_j}\right)\right\|^2_2\\
\notag&-\frac{r-a-b}{abr}\left\|\mathcal{A}\left(bZ^{(r)}_{T_i}-aZ^{(r)}_{S_j}\right)\right\|^2_2\bigg]\\
&\leq\rho_{a,b}(t)t[t-(2-t)\delta_{tr}]\|Z^{(r)}\|^2_F,
\end{align}
and for $1\leq t<4/3$,
\begin{align}\label{eq.26}
\notag&\rho_{a,b}(t)\sum_k\tau_k\bigg[\left\|\mathcal{A}\left(Z^{(r)}+(t-1)W_k\right)\right\|^2_2\\
\notag&-\left\|(t-1)\mathcal{A}\left(Z^{(r)}+W_k\right)\right\|^2_2\bigg]\\
\notag&\leq\rho_{a,b}(t)\bigg\{\bigg[t(2-t)-(t^2-2t+2)\delta_{tr}\bigg]\|Z^{(r)}\|^2_F\\
&-2r\alpha^2\delta_{tr}(t-1)\bigg\},
\end{align}
where
$$\rho_{a,b}(t)=(a+b)^2-2ab(4-t).$$
\end{lemma}

\noindent \textbf{Proof.}
We first consider the case of $0<t<1$. As $tr$ equals to even, we can fix $a=b=tr/2$; And as $tr$ equals to odd, we can set $a=b+1=(tr+1)/2$; For both cases, one can easily prove that $\rho_{a,b}(t)<0$. Since $Z^{(r)}_{T_i},~Z^{(r)}_{S_j}$ are $a$-rank and $b$-rank, respectively, by utilizing $tr$-order RIP, we get
\begin{align}\label{eq.27}
\notag&\frac{\rho_{a,b}(t)}{(^r_a)(^{r-a}_b)}\sum_{T_i\bigcap S_j=\emptyset}\bigg[\left\|\mathcal{A}\left(Z^{(r)}_{T_i}+Z^{(r)}_{S_j}\right)\right\|^2_2\\
\notag&-\frac{r-a-b}{abr}\left\|\mathcal{A}\left(bZ^{(r)}_{T_i}-aZ^{(r)}_{S_j}\right)\right\|^2_2\bigg]\\
\notag\leq&\frac{\rho_{a,b}(t)}{(^r_a)(^{r-a}_b)}\sum_{T_i\bigcap S_j=\emptyset}\bigg[(1-\delta_{tr})\left\|Z^{(r)}_{T_i}+Z^{(r)}_{S_j}\right\|^2_F\\
\notag&-\frac{r-a-b}{abr}(1+\delta_{tr})\left\|bZ^{(r)}_{T_i}-aZ^{(r)}_{S_j}\right\|^2_F\bigg]\\
\notag=&\frac{\rho_{a,b}(t)}{(^r_a)(^{r-a}_b)}\bigg\{(1-\delta_{tr})\bigg[(^{r-a}_b)\sum_{i\in A}\|Z^{(r)}_{T_i}\|^2_F\\
\notag&+(^{r-b}_a)\sum_{j\in B}\|Z^{(r)}_{S_j}\|^2_F\bigg]\\
\notag&-\frac{1-t}{ab}(1+\delta_{tr})\bigg[b^2(^{r-a}_b)\sum_{i\in A}\|Z^{(r)}_{T_i}\|^2_F\\
\notag&+a^2(^{r-b}_a)\sum_{j\in B}\|Z^{(r)}_{S_j}\|^2_F\bigg]\bigg\}\\
\notag=&\frac{\rho_{a,b}(t)}{(^r_a)(^{r-a}_b)}\bigg\{(1-\delta_{tr})\bigg[(^{r-a}_b)(^{r-1}_{a-1})
+(^{r-b}_a)(^{r-1}_{b-1})\bigg]\|Z^{(r)}\|^2_F\\
\notag&-(1+\delta_{tr})\frac{1-t}{ab}\left[b^2(^{r-a}_b)(^{r-1}_{a-1})
+a^2(^{r-b}_a)(^{r-1}_{b-1})\right]\|Z^{(r)}\|^2_F\bigg\}\\
=&\rho_{a,b}(t)t[t-(2-t)\delta_{tr}]\|Z^{(r)}\|^2_F,
\end{align}
where we made use of Lemma $\ref{le.1}$ to the second equality.

Next, we discuss the case of $1\leq t<4/3$.

Observe that $Z^{(r)},~W_k$ are $r$-rank and $(t-1)r$-rank, respectively. Under the assumption of $\rho_{a,b}(t)<0$, combining with $tr$-order RIP, we get
\begin{align}\label{eq.28}
\notag&\rho_{a,b}(t)\sum_k\tau_k\bigg[\left\|\mathcal{A}\left(Z^{(r)}+(t-1)W_k\right)\right\|^2_F\\
\notag&-\left\|(t-1)\mathcal{A}\left(Z^{(r)}+W_k\right)\right\|^2_F\bigg]\\
\notag\leq&\rho_{a,b}(t)\sum_k\tau_k\bigg[(1-\delta_{tr})\left\|Z^{(r)}+(t-1)W_k\right\|^2_F\\
\notag&-(t-1)^2(1+\delta_{tr})\left\|Z^{(r)}+W_k\right\|^2_F\bigg]\\
\notag=&\rho_{a,b}(t)\sum_k\tau_k\bigg\{(1-\delta_{tr})\bigg[\|Z^{(r)}\|^2_F
+(t-1)^2\|W_k\|^2_F\bigg]\\
\notag&-(t-1)^2(1+\delta_{tr})\bigg(\|Z^{(r)}\|^2_F+\|W_k\|^2_F\bigg)\bigg\}\\
\notag=&\rho_{a,b}(t)\bigg\{\bigg[(1-\delta_{tr})-(t-1)^2(1+\delta_{tr})\bigg]\|Z^{(r)}\|^2_F\\
\notag&-2\delta_{tr}(t-1)^2\sum_k\tau_k\|W_k\|^2_F\bigg\}\\
\notag&\leq\rho_{a,b}(t)\bigg\{\bigg[t(2-t)-(t^2-2t+2)\delta_{tr}\bigg]\|Z^{(r)}\|^2_F\\
&-2r\alpha^2\delta_{tr}(t-1)\bigg\},
\end{align}
where the first equality follows from the fact that $\left<Z^{(r)},W_k\right>=0$, and for the last inequality, we used the inequality (\ref{eq.15}).

\qed

As we described in the Introduction part, Cai and Zhang \cite{Cai and Zhang} established the sharp sufficient conditions to ensure the recovery of low-rank matrices via nuclear norm minimization. Their main results are stated as follows.
\begin{theorem}\label{th.4}
(Theorem $3.7$ in \cite{Cai and Zhang}) Consider the affine rank minimization problem $b=\mathcal{A}X+z$ with $\|z\|_2\leq \epsilon$. Let $X_*$ be the minimizer of $\arg\min\{\|X\|_*:~AX-z\in\mathcal{B}\}$ with $\mathcal{B}=\{z:~\|z\|_2\leq\eta\}$ for some $\eta\geq\epsilon$. If
$\delta_{r}<1/3$ with $r\geq 2$, then
\begin{align}
\notag&\|X_*-X\|_F\leq\frac{\sqrt{2(1+\delta_r)}}{1-3\delta_r}(\epsilon+\eta)\\
\notag&+\frac{2\sqrt{2}}{\sqrt{r}}\left[\frac{\sqrt{2}}{2}
+\frac{2\delta_r+\sqrt{(1-3\delta_r)\delta_r}}{1-3\delta_r}\right]\cdot\|X-X^{(r)}\|_*.
\end{align}
\end{theorem}

\begin{theorem}\label{th.5}
(Theorem $3.8$ in \cite{Cai and Zhang}) Consider the affine rank minimization problem $b=\mathcal{A}X+z$ with $\|\mathcal{A}^*(z)\|\leq \epsilon$. Let $X_*$ be the minimizer of $\arg\min\{\|X\|_*:~AX-z\in\mathcal{B}\}$ with $\mathcal{B}=\{z:~\|\mathcal{A}^*(z)\|\leq\eta\}$ for some $\eta\geq\epsilon$. If
$\delta_{r}<1/3$ with $r\geq 2$, then
\begin{align}
\notag&\|X_*-X\|_F\leq\frac{\sqrt{2r}}{1-3\delta_r}(\epsilon+\eta)\\
\notag&+\frac{2\sqrt{2}}{\sqrt{r}}\left[\frac{\sqrt{2}}{2}
+\frac{2\delta_r+\sqrt{(1-3\delta_r)\delta_r}}{1-3\delta_r}\right]\cdot\|X-X^{(r)}\|_*.
\end{align}
\end{theorem}

\section{Main results}
\label{sec.3}

\begin{theorem}\label{th.1}
Consider rank minimization problem $b=\mathcal{A}X+z$ with $\|z\|_2\leq \epsilon$. If
$\delta_{tr}<t/(4-t)$ with $0<t<4/3$, then the solution $X^*$ to the nuclear norm minimization problem (\ref{eq.2}) fulfils
\begin{align}\label{eq.29}
\|X-X^*\|_F\leq C_1\epsilon+C_2\|X-X^{(r)}\|_*,
\end{align}
where
\begin{align}\label{eq.30}
C_1=\frac{2\sqrt{2(1+\delta_{tr})}\kappa}{\frac{t}{4-t}-\delta_{tr}},
\end{align}
and
\begin{align}\label{eq.31}
C_2=\frac{2\sqrt{2}}{\sqrt{r}}\left\{\frac{1}{4}+\frac{2\delta_{tr}
+\sqrt{\delta_{tr}(4-t)\left(\frac{t}{4-t}-\delta_{tr}\right)}}{\frac{t}{4-t}-\delta_{tr}}\right\}
\end{align}
with
\begin{align}
\notag \kappa=\max\left\{\frac{t}{4-t},~\frac{\sqrt{t}}{4-t}\right\}.
\end{align}
Similarly, Consider rank minimization problem $b=\mathcal{A}X+z$ with $z$ such that $\|\mathcal{A}^*(z)\|\leq\epsilon$. If
$\delta_{tr}<t/(4-t)$ with $0<t<4/3$, then the solution $X^{\circ}$ to the nuclear norm minimization problem $\min_{X}~\|X\|_*~\mbox{s.t.}~\|\mathcal{A}^*(z)\|\leq\epsilon$ fulfils
\begin{align}\label{eq.32}
\|X-X^{\circ}\|_F\leq C_3\epsilon+C_4\|X-X^{(r)}\|_*,
\end{align}
where
\begin{align}\label{eq.33}
C_3=\frac{2\sqrt{2r}\kappa}{\frac{t}{4-t}-\delta_{tr}},
\end{align}
and $C_4=C_2$.
\end{theorem}

\begin{remark}
As $t=1$, the upper bound $\delta_r<1/3$ is coincident with Theorems $\ref{th.4}$ and $\ref{th.5}$ of \cite{Cai and Zhang}. Furthermore, the upper bounds of error estimates $\|X-X^*\|_F~(\|X-X^{\circ}\|_F)$ are smaller than the results of \cite{Cai and Zhang}. In theory, the recovered precision is given by our results is higher than that of theirs.
\end{remark}

\begin{corollary}
Assume that $X\in\mathbb{R}^{m\times n}$ is a $r$-rank matrix. Let $b=\mathcal{A}X$. If
\begin{align}\label{eq.34}
\delta_{tr}<t/(4-t)
\end{align}
for $0<t<4/3$, then the solution $X^*$ to the nuclear norm minimization problem (\ref{eq.2}) in the noiseless case (i.e., $\epsilon=0$) reconstructs $X$ exactly.
\end{corollary}

\begin{remark}
As $t=1$, the upper bound $\delta_r<1/3$ is the same as Theorem $3.5$ of \cite{Cai and Zhang}.
\end{remark}

The Gaussian noise situation is of special interest in statistics and image processing. Note that the Gaussian random variables are essentially bounded. The results given in Theorem $\ref{th.1}$ regarding the bounded noise situation are immediately applied to the Gaussian noise situation, which employs the similar discussion as that in \cite{Lin et al 2013}.

\begin{theorem}\label{th.3}
Assume that the low-rank recovery model $b=\mathcal{A}X+z$ with $z\sim N_q(0,\sigma^2I)$.
$\delta_{tr}<t/(4-t)$ for some $0<t<4/3$. Let $X^*$ represent the minimizer of $\min_{X}~\|X\|_*~\mbox{s.t.}~\|z\|_2\leq\sigma\sqrt{q+2\sqrt{q\log q}}$ and let $X^{\circ}$ be the minimizer of $\min_{X}~\|X\|_*~\mbox{s.t.}~\|\mathcal{A}^*(z)\|\leq2\sigma\sqrt{\log n}$. We have
with probability at least $1-1/q$,
\begin{align}
\notag&\|X-X^*\|_F\leq \frac{2\sqrt{2(1+\delta_{tr})}\kappa}{\frac{t}{4-t}-\delta_{tr}}\sigma\sqrt{q+2\sqrt{q\log q}}\\
\notag&+2\sqrt{2}\left\{\frac{1}{4}+\frac{2\delta_{tr}
+\sqrt{\delta_{tr}(4-t)\left(\frac{t}{4-t}-\delta_{tr}\right)}}{\frac{t}{4-t}-\delta_{tr}}\right\}\frac{\|X-X^{(r)}\|_*}{\sqrt{r}},
\end{align}
and probability at least $1-1/\sqrt{\pi\log n}$,
\begin{align}
\notag&\|X-X^{\circ}\|_F\leq \frac{4\sqrt{2r}\kappa}{\frac{t}{4-t}-\delta_{tr}}\sigma\sqrt{\log n}\\
\notag&+2\sqrt{2}\left\{\frac{1}{4}+\frac{2\delta_{tr}
+\sqrt{\delta_{tr}(4-t)\left(\frac{t}{4-t}-\delta_{tr}\right)}}{\frac{t}{4-t}-\delta_{tr}}\right\}\frac{\|X-X^{(r)}\|_*}{\sqrt{r}},
\end{align}
where
$\kappa$ is defined in Theorem $\ref{th.1}$.
\end{theorem}

\begin{theorem}\label{th.2}
Let $1\leq r\leq m/2$. There is a linear map $\mathcal{A}:~\mathbb{R}^{m\times m}\to\mathbb{R}^{q}$ with $\delta_{tr}<t/(4-t)+\varepsilon$ with $0<t<4/3,~\varepsilon>0$ such that for some $r$-rank matrices $Y_1,~Y_2\in\mathbb{R}^{m\times m}$ with $Y_1\neq Y_2$, $\mathcal{A}Y_1=\mathcal{A}Y_2$. Hence, there don't exist any approach to exactly reconstruct all $r$-rank matrices $X$ based on $(\mathcal{A},z)$.
\end{theorem}

\begin{remark}
Theorems $\ref{th.1}$ and $\ref{th.2}$ jointly indicate the condition $\delta_{tr}<t/(4-t)$ with $0<t<4/3$ is sharp.
\end{remark}

\section{Numerical experiments}
\label{sec.4}

\subsection{Solution algorithm}

In this section, we carry out some numerical experiments to verify our theoretical results. In order to solve the nuclear norm minimization model (\ref{eq.2}), we will utilize alternating direction method of multipliers (abbreviated as ADMM), which is usually applied in sparse signal recovery and low-rank matrix reconstruction, see references \cite{Lu C et al 2018} \cite{Wen F et al 2017} \cite{Wang W et al 2017}. The constrained optimization problem (\ref{eq.2}) could be converted into the unconstrained optimization problem as follows:
\begin{align}\label{eq.47}
\min_{\hat{X}}\|\hat{X}\|_*+\frac{\lambda}{2}\|A\mbox{vec}(\hat{X})-b\|^2_2,
\end{align}
where $\lambda$ is a regularization parameter, and $\mbox{vec}(\hat{X})$ denotes the vectorization of $\hat{X}$. Then, we employ ADMM technique to solve the problem (\ref{eq.47}). In particular, introducing a new variable $V\in\mathbb{R}^{m\times n}$, the problem (\ref{eq.47}) could be equivalently transformed into
\begin{align}\label{eq.48}
\min_{\hat{X}}\|V\|_*+\frac{\lambda}{2}\|A\mbox{vec}(\hat{X})-b\|^2_2~\mbox{s.t.}~\hat{X}=V.
\end{align}
The augmented Lagrangian function is
\begin{align}\label{eq.49}
\notag L(\hat{X},V,Y)=&\|V\|_*+\frac{\lambda}{2}\|A\mbox{vec}(\hat{X})-b\|^2_2
+\left<Y,\hat{X}-V\right>\\
&+\frac{\mu}{2}\|\hat{X}-V\|^2_F,
\end{align}
where $Y\in\mathbb{R}^{m\times n}$ is the dual variable, and $\mu$ is the penalty parameter associating to augmented Lagrangian function. Then, applying ADMM to (\ref{eq.49}), we could obtain the following iterations:
\begin{align}\label{eq.50}
\notag\hat{X}^{k+1}&=\arg\min_{\hat{X}\in\mathbb{R}^{m\times n}}\frac{\lambda}{2}\|A\mbox{vec}(\hat{X})-b\|^2_2\\
\notag&+\frac{\mu}{2}\|\hat{X}-V^k+\frac{Y^k}{\mu}\|^2_F,\\
\notag V^{k+1}&=\arg\min_{V\in\mathbb{R}^{m\times n}}\|V\|_*+\frac{\mu}{2}\|\hat{X}-V^k+\frac{Y^k}{\mu}\|^2_F,\\
Y^{k+1}&=Y^k+\hat{X}^{k+1}-V^{k+1}.
\end{align}

In the experiment, the $r$-rank matrix $X\in\mathbb{R}^{m\times n}$ is generated by $X=P*Q$, where $P\in\mathbb{R}^{m\times r}$ and $Q\in\mathbb{R}^{r\times n}$. We produce the measurement matrix $A\in\mathbb{R}^{q\times mn}$ with its elements being i.i.d. zero mean and $1/q$ Gaussian random variables. In all experiments, we take $m=n=50$, and $r=0.2*m$. On the premise that $A$ and $X$ are known, the measurement $b$ is produced by $b=A\mbox{vec}(\hat{X})+\epsilon*z$, where the entries of $z$ follow zero mean and 0.05 standard variation Gaussian distribution, and $\epsilon$ represents the noise level whose range of value is 0, 0.05 and 0.1. In all experiments, we report the average result over 50 independent tests.

\subsection{Algorithm convergence}

\begin{figure}[h]
        \centering
		\includegraphics[scale=0.2]{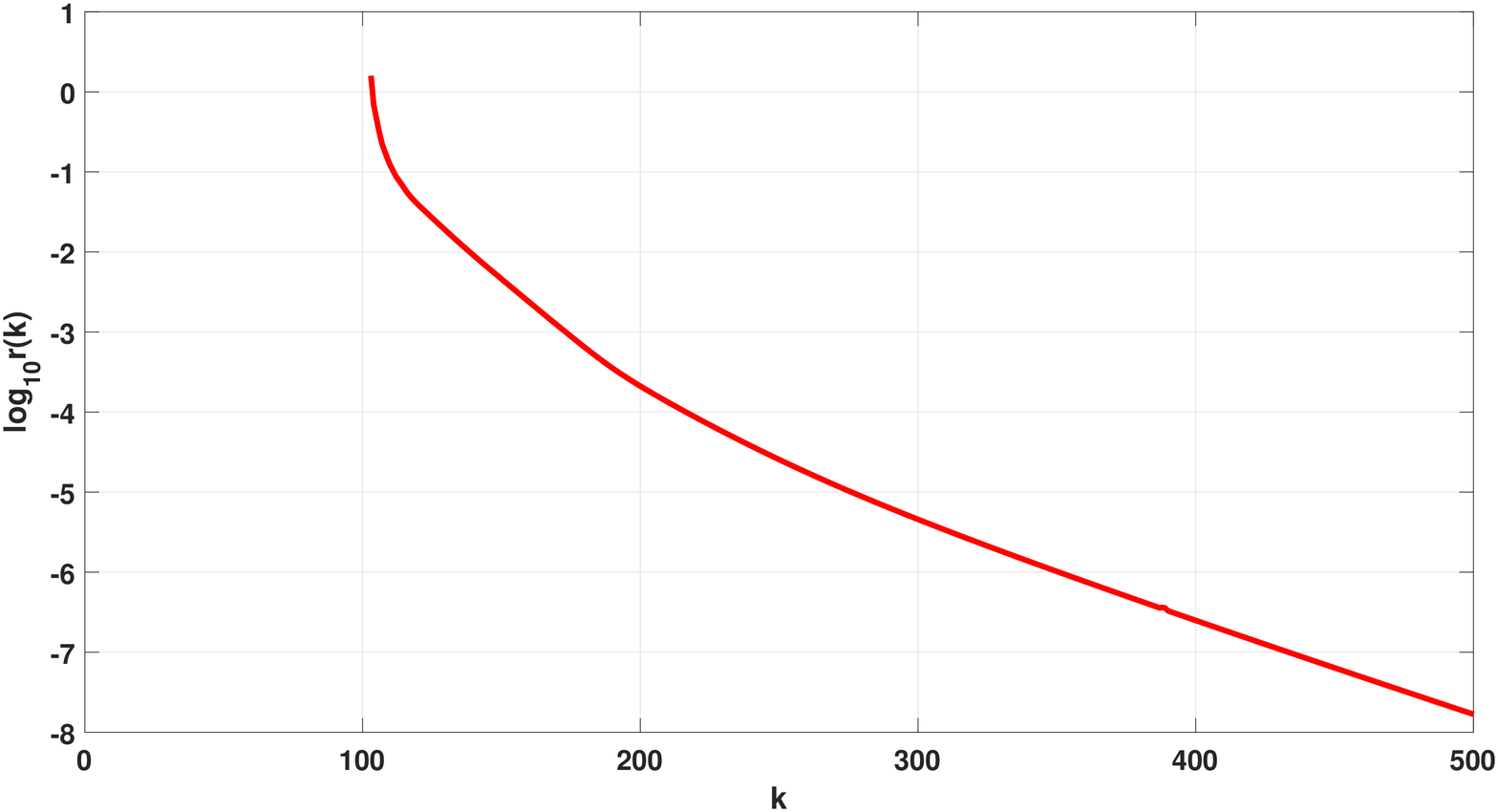}
\caption{Algorithm convergence for problem (\ref{eq.2})}\label{fig.1}
\end{figure}

Fig. \ref{fig.1} shows the result about algorithm convergence for solving the problem (\ref{eq.2}). It is observed that the relative neighboring iteration error ($r(k)=\|X^{k+1}-X^k\|_F/\|X^k\|_F$) decreases with the increase of iteration times $k$. When the number of iterations $k$ exceeds 210, it tends to become less than $10^{-4}$.

\subsection{Comparison of error bounds}

\begin{figure}[htbp]
\begin{center}
\subfigure[]{\includegraphics[width=0.40\textwidth]{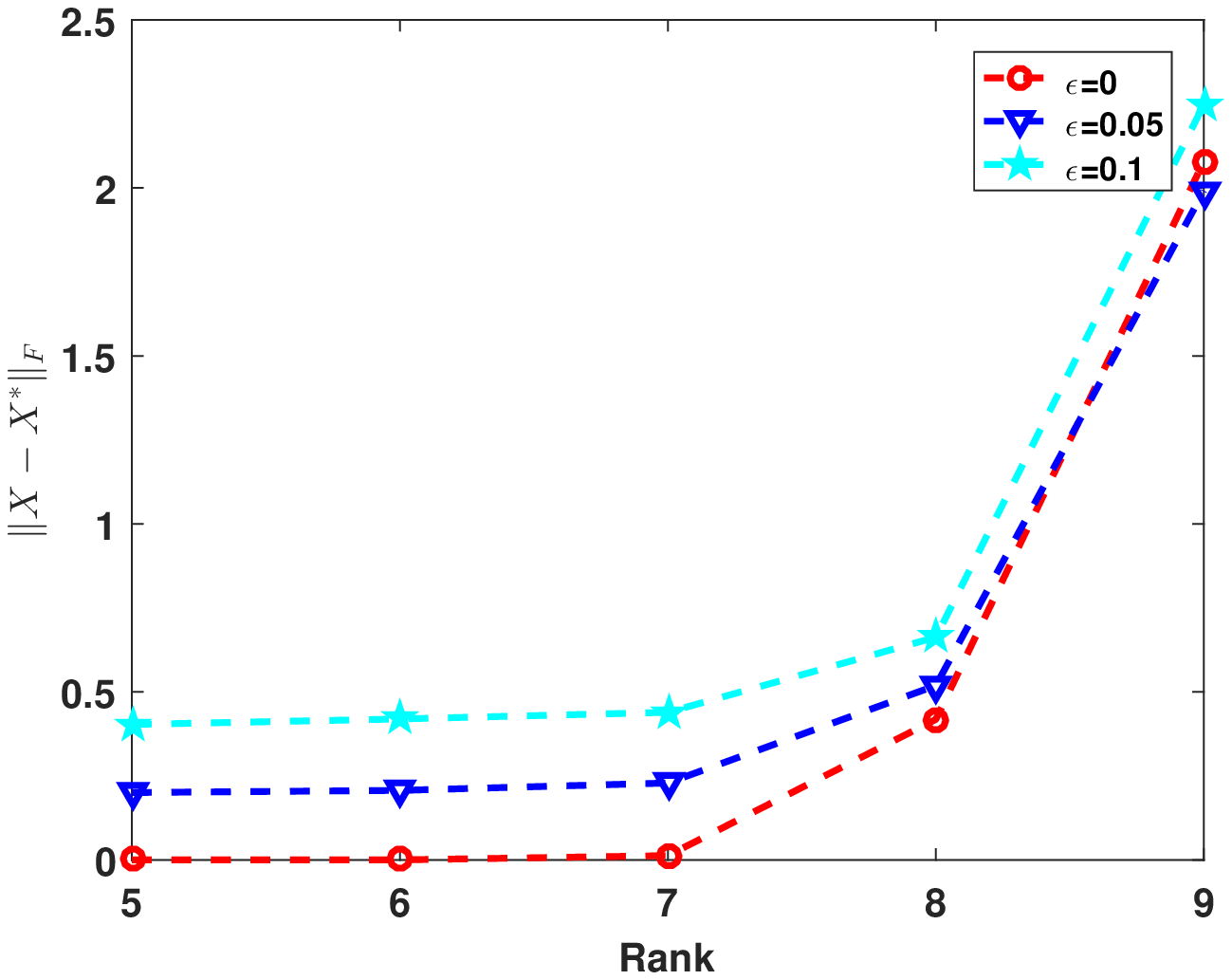}}
\subfigure[]{\includegraphics[width=0.40\textwidth]{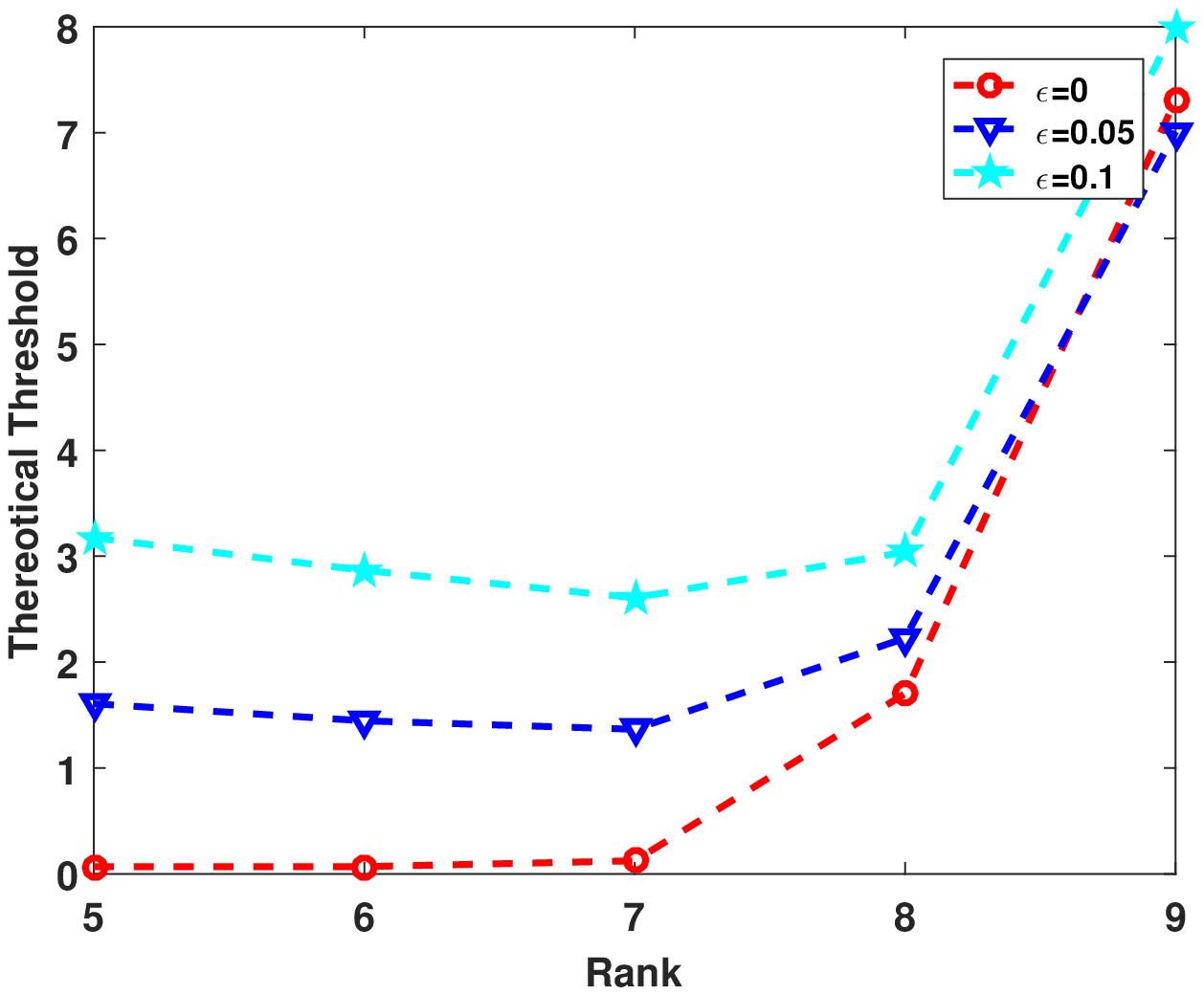}}
\caption{Bounded noise, (a) $\|X-X^*\|_F$ versus rank $r$, (b) the theoretical error bound given by (\ref{eq.29}) for $t=1$ and $\delta_r=0.05$.}\label{fig.2}
\end{center}
\vspace*{-14pt}
\end{figure}

\begin{figure}[htbp]
\begin{center}
\subfigure[]{\includegraphics[width=0.40\textwidth]{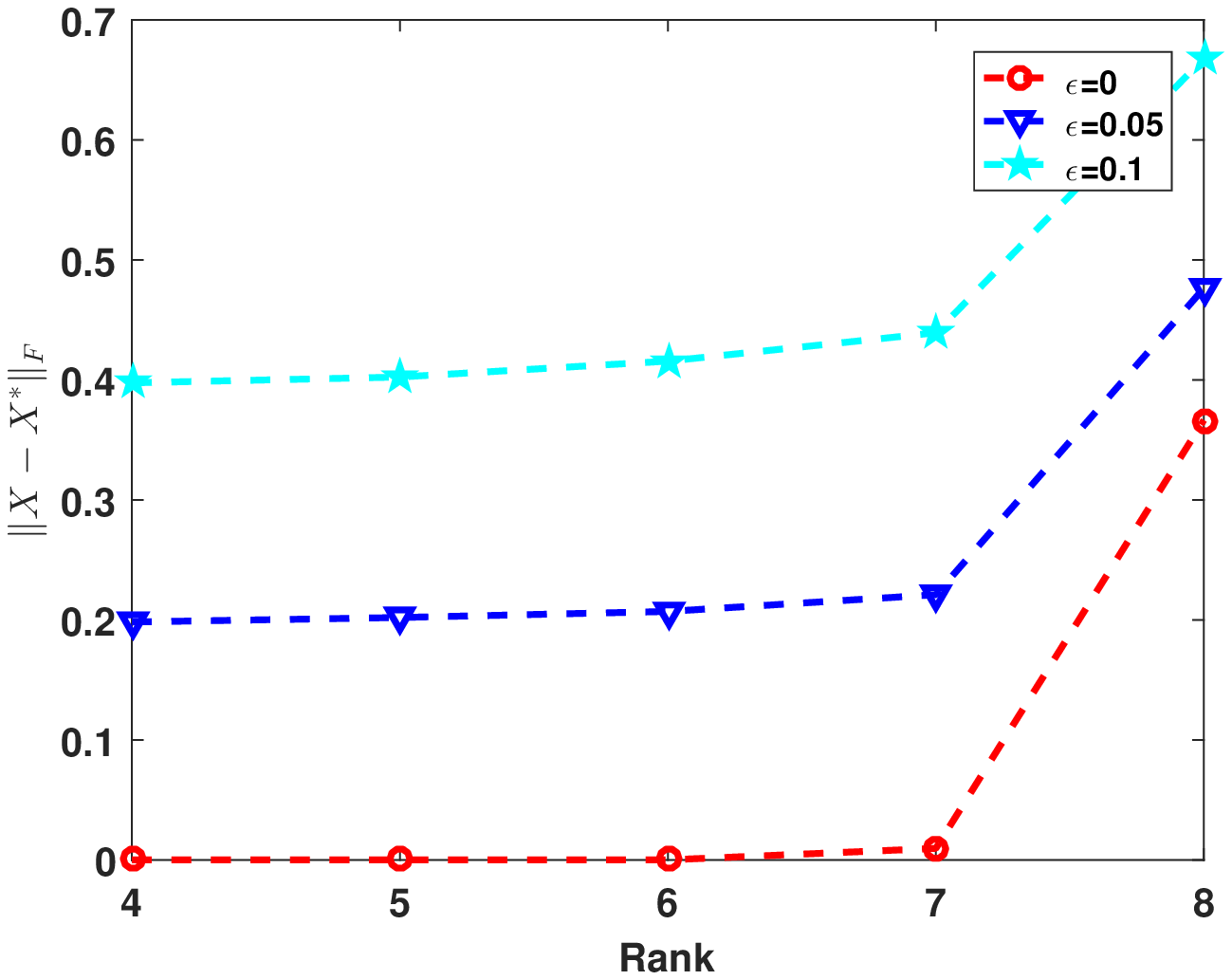}}
\subfigure[]{\includegraphics[width=0.40\textwidth]{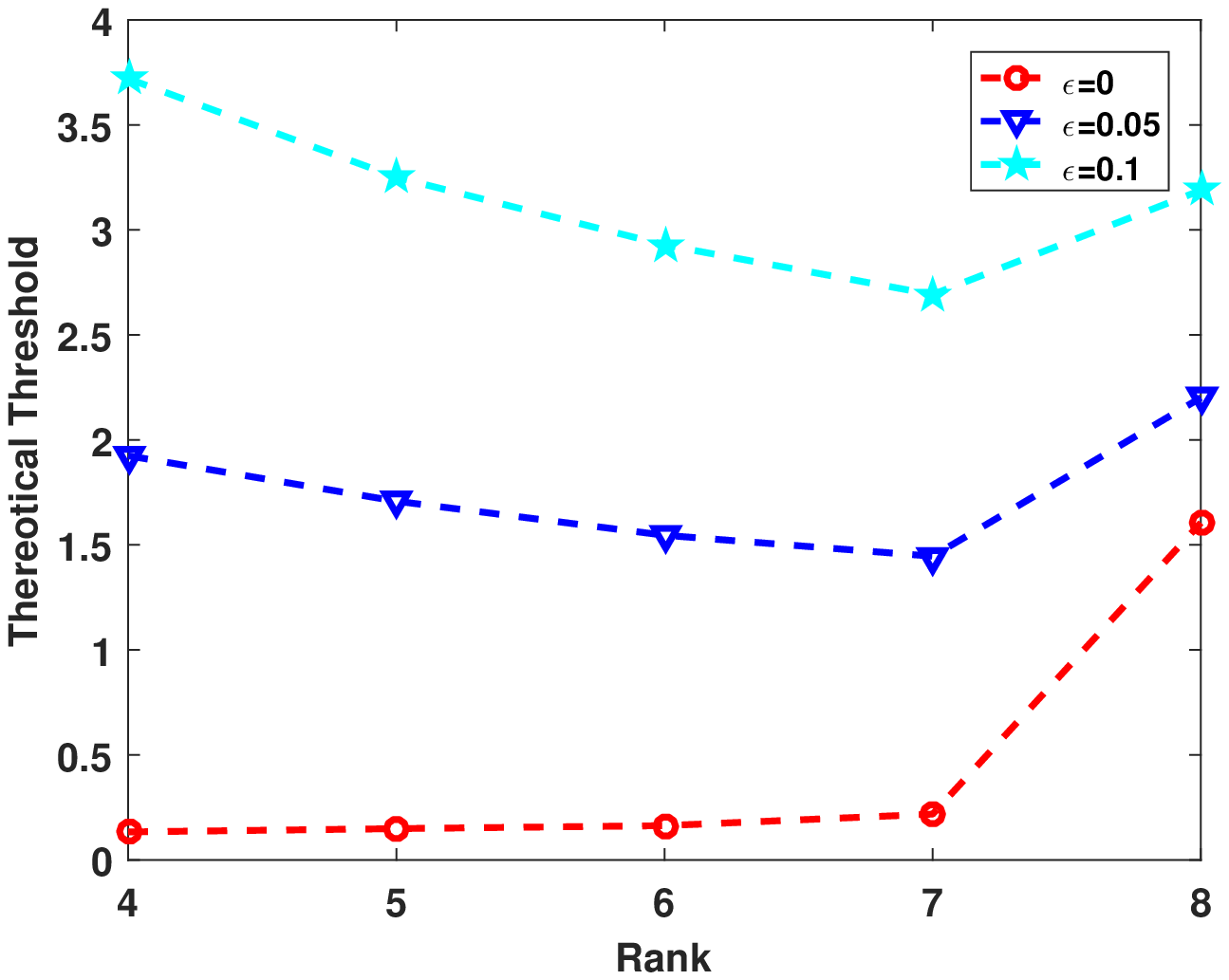}}
\caption{Dantzig selector noise, (a) $\|X-X^*\|_F$ versus rank $r$, (b) the theoretical error bound given by (\ref{eq.32}) for $t=1$ and $\delta_r=0.05$.}\label{fig.3}
\end{center}
\vspace*{-14pt}
\end{figure}

In Fig. \ref{fig.2}(a) $\|X-X^*\|_F$ is plotted versus the rank $r$ for different noise level $\epsilon=0,~0.05,~0.1$. In Fig. \ref{fig.2}(b) the relevant theoretical error bound determined by (\ref{eq.29}) is plotted with $t=1$ and $\delta_r=0.05$. One can easily see that $\|X-X^*\|_F$ is lower than the theoretical error bound. Fig. \ref{fig.3}(a) and (b) present $\|X-X^*\|_F$ and the corresponding theoretical error bound  defined by (\ref{eq.32}).

\subsection{Results of different measurement matrices}

\begin{figure}[htbp]
\begin{center}
\subfigure[]{\includegraphics[width=0.40\textwidth]{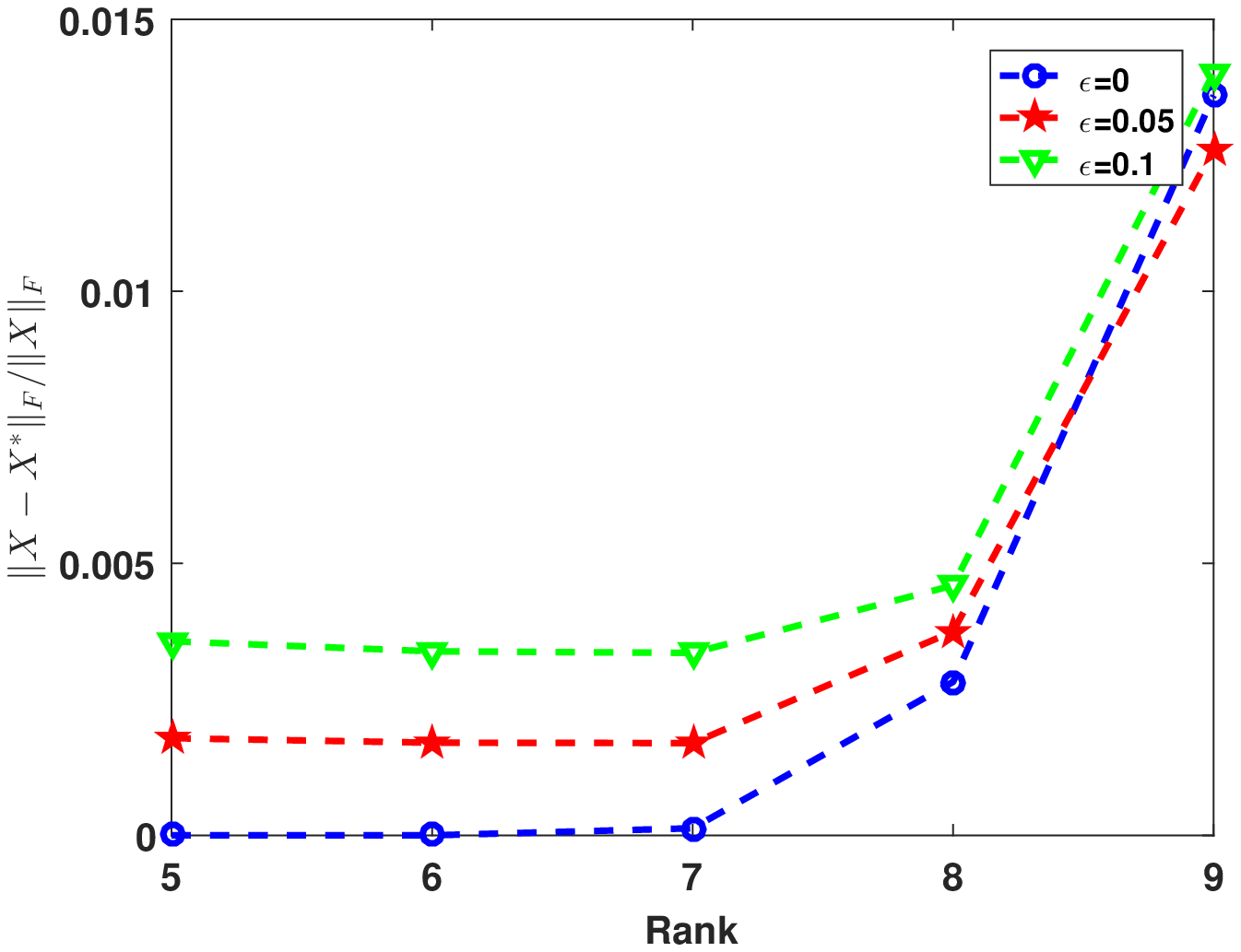}}
\subfigure[]{\includegraphics[width=0.40\textwidth]{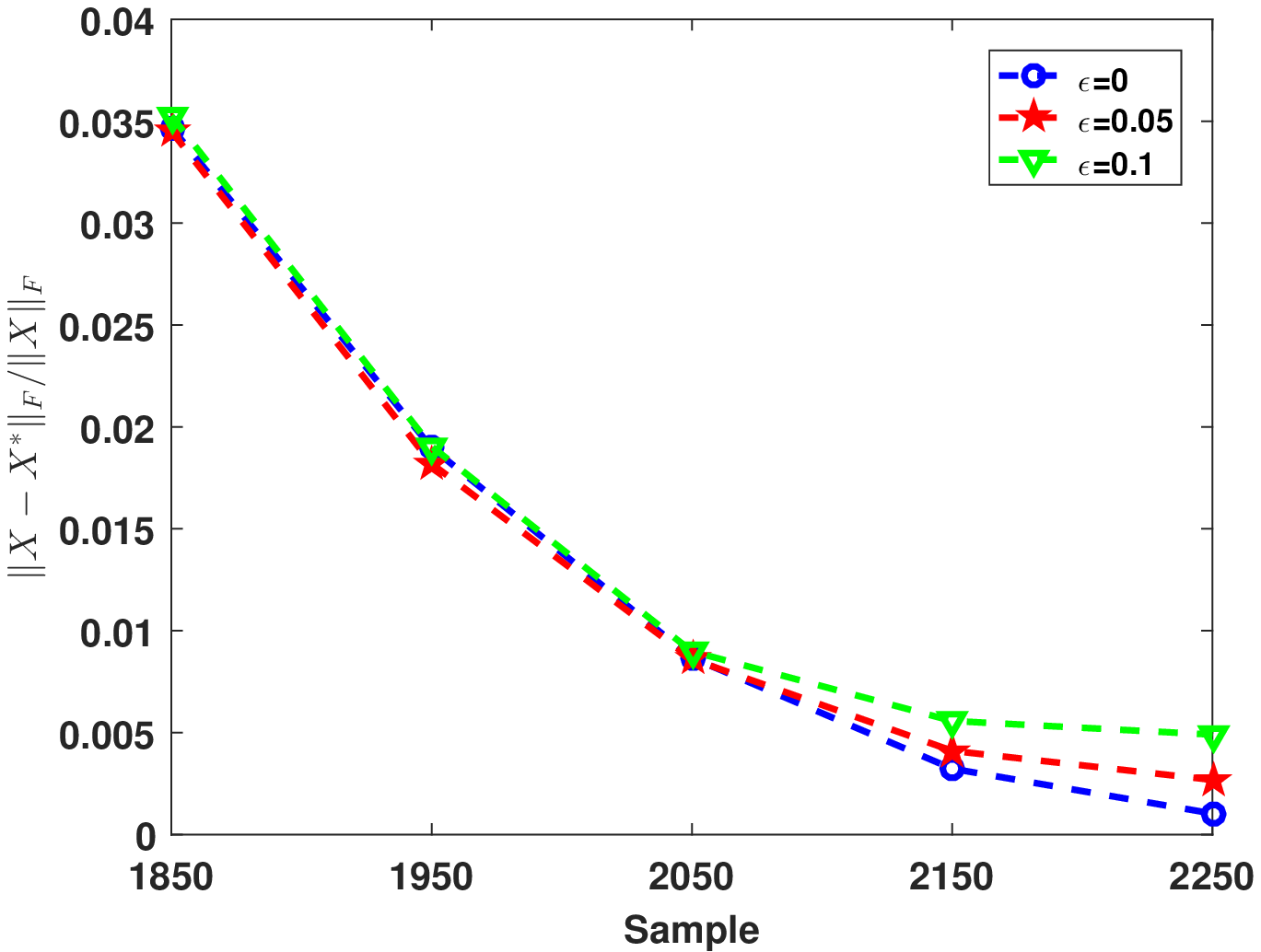}}
\caption{Gaussian measurement matrix, (a) relative error versus rank $r$, (b) relative error versus number of measurement $q$.}\label{fig.4}
\end{center}
\vspace*{-14pt}
\end{figure}

Fig. \ref{fig.4}(a) plots the relationship between the relative error $\|X-X^*\|_F/\|X\|_F$ and the rank $r$ for Gaussian measurement matrix. Fig. \ref{fig.4}(b) plots the relation the relative error and the number of measurement $q$. It is easy to see that a decreasing rank $r$ or an increasing number of measurement $q$ leads to a better performance of the model (\ref{eq.2}). Furthermore, the smaller the noise level, the better the model reconstruction effect.

\begin{figure}[htbp]
\begin{center}
\subfigure[]{\includegraphics[width=0.40\textwidth]{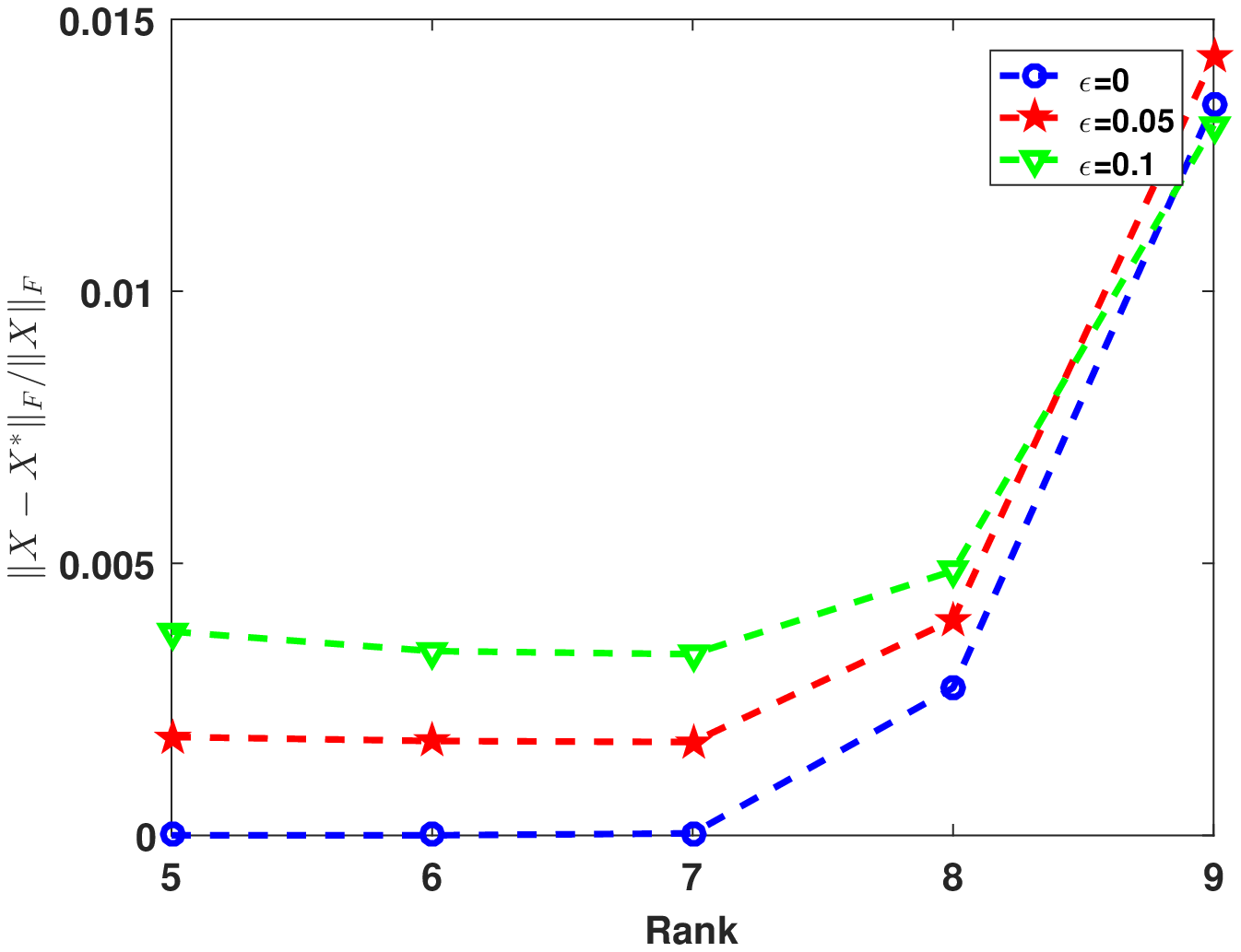}}
\subfigure[]{\includegraphics[width=0.40\textwidth]{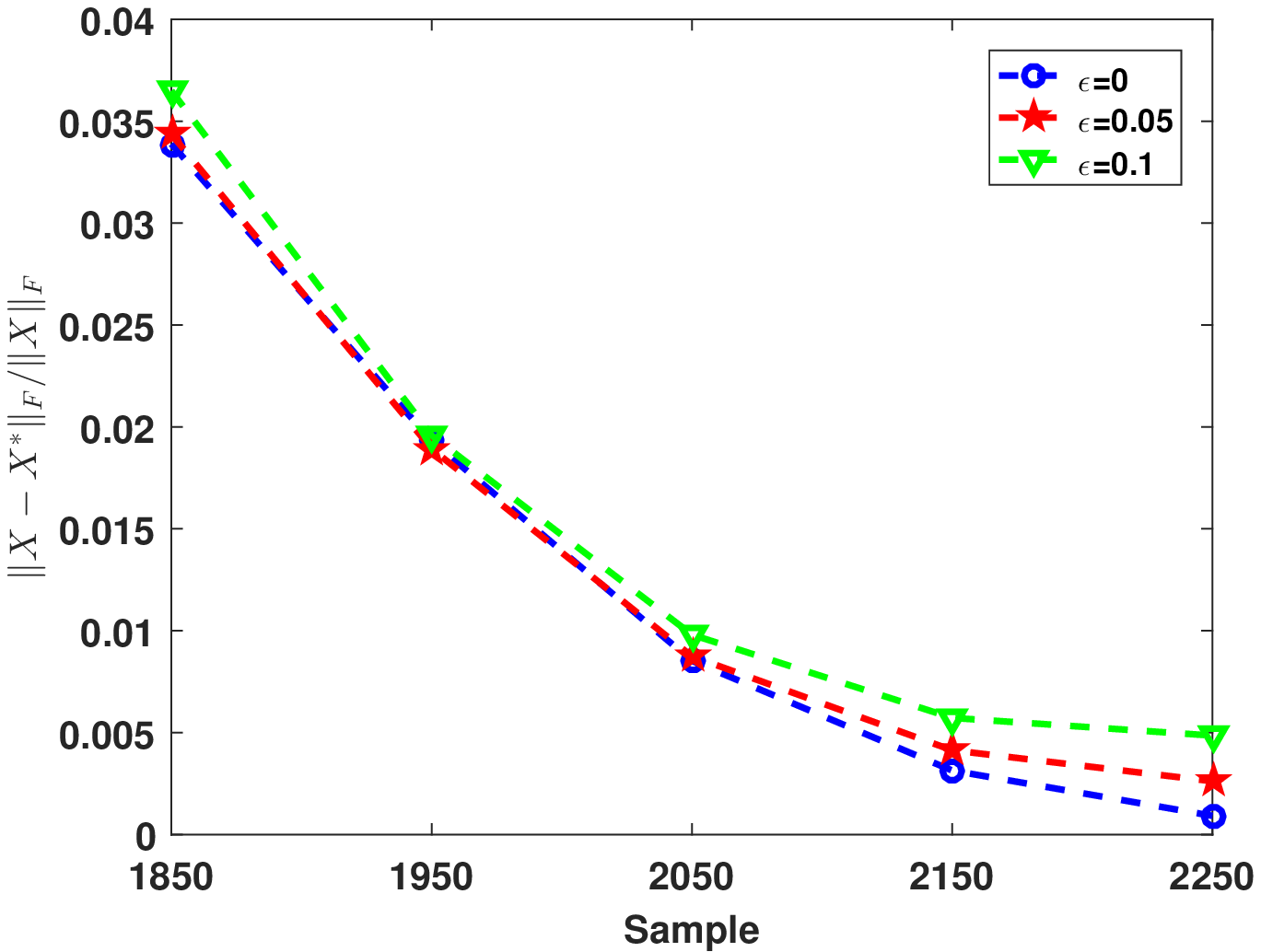}}
\caption{Bernoulli measurement matrix, (a) relative error versus rank $r$, (b) relative error versus number of measurement $q$.}\label{fig.5}
\end{center}
\vspace*{-14pt}
\end{figure}

\begin{figure}[htbp]
\begin{center}
\subfigure[]{\includegraphics[width=0.40\textwidth]{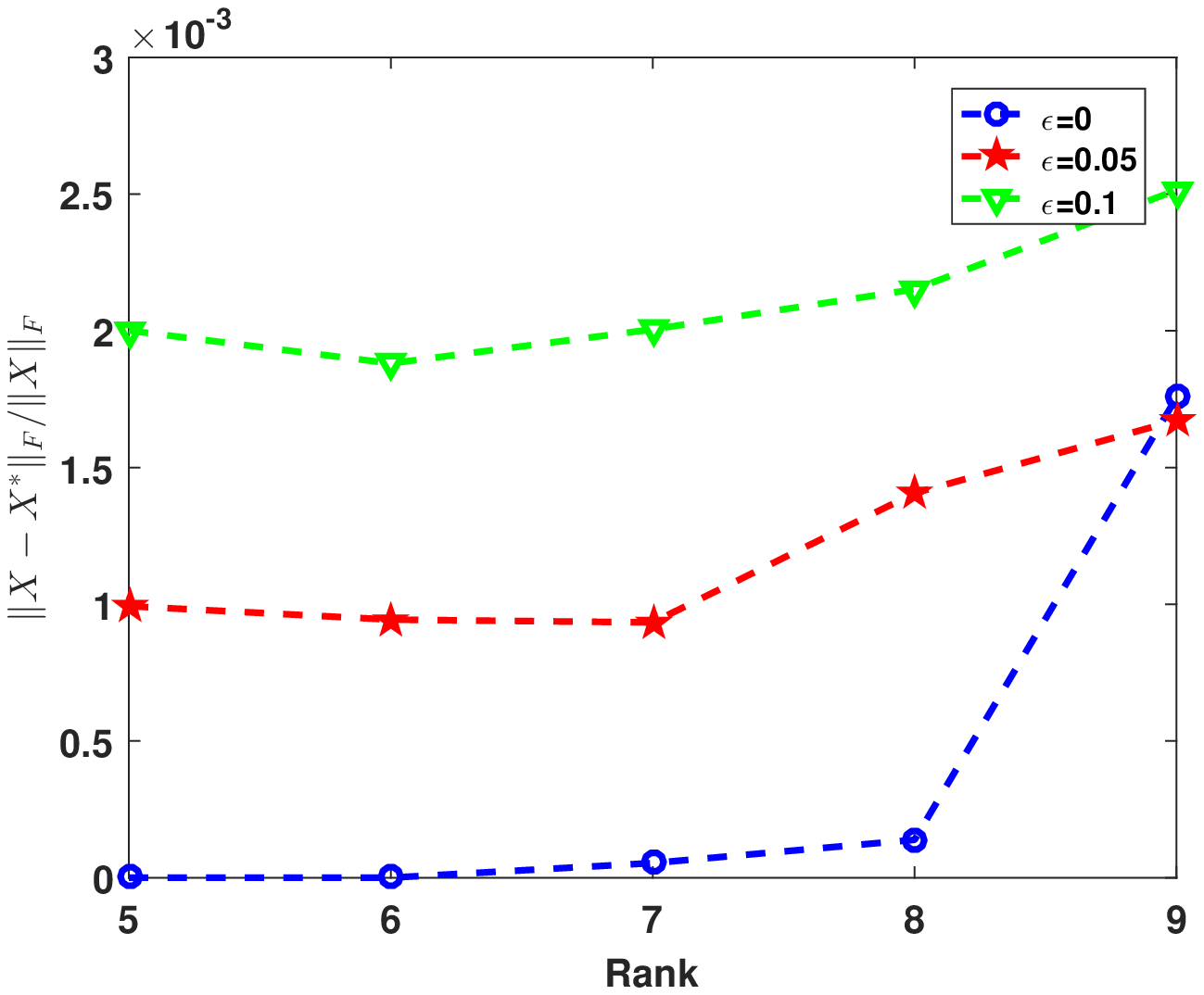}}
\subfigure[]{\includegraphics[width=0.40\textwidth]{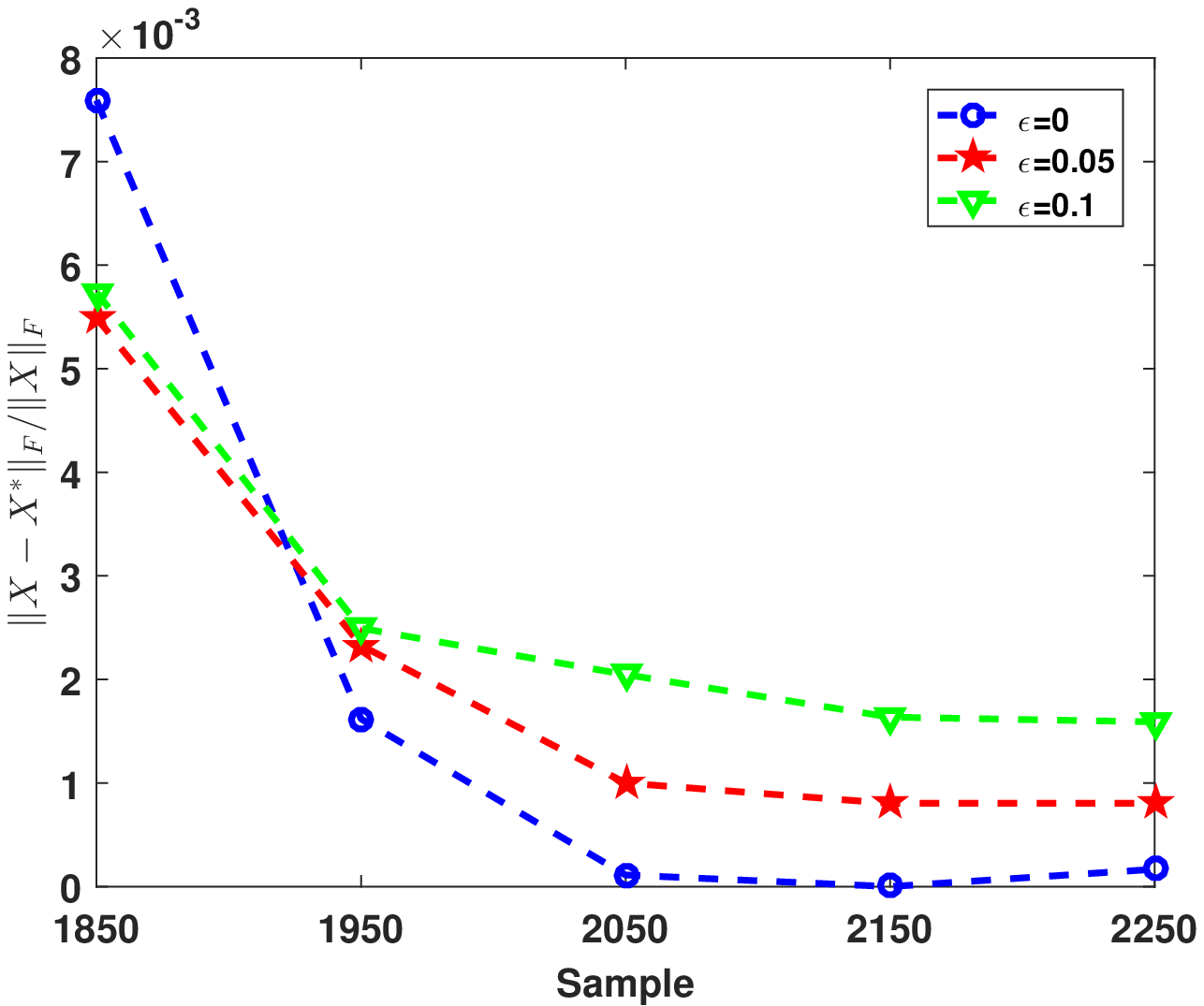}}
\caption{Partial Fourier measurement matrix, (a) relative error versus rank $r$, (b) relative error versus number of measurement $q$.}\label{fig.6}
\end{center}
\vspace*{-14pt}
\end{figure}

In Figs. \ref{fig.5} and \ref{fig.6}, the relative errors are plotted respectively for Bernoulli measurement matrix and Partial Fourier measurement matrix. It is observed from Figs. \ref{fig.4}, \ref{fig.5} and \ref{fig.6} that the reconstruction performance of the nuclear norm minimization method (\ref{eq.2}) is the best when the measurement matrix is a partial Fourier matrix.

\section{Proofs of main results}
\label{sec.5}

With above preparation, we present the proof of main results.

\noindent \textbf{Proof of Theorem $\ref{th.1}$.} By the definition of $\alpha$ and notice that the rank of $Z^{(r)}$ is at most $r$, we get
\begin{align}\label{eq.35}
\notag\alpha^2&=\frac{\|Z^{(r)}\|^2_*+4\|Z^{(r)}\|_*\|X-X^{(r)}\|_*+4\|X-X^{(r)}\|^2_*}{r^2}\\
&\leq\frac{\|Z^{(r)}\|^2_F}{r}+\frac{4\|Z^{(r)}\|_F\|X-X^{(r)}\|_*}{r\sqrt{r}}
+\frac{4\|X-X^{(r)}\|^2_*}{r^2},
\end{align}
where in the last step, we used the fact that for any $X\in\mathbb{R}^{m\times n}~(m\leq n)$ and $p\in(0,1],$
\begin{align}\label{eq.36}
m^{\frac{1}{p}-\frac{1}{2}}\|X\|_F\geq\|X\|_p
\end{align}
with $\|X\|_p=(\sum_i\sigma^p_i(X))^{1/p}$. Additionally, due to the feasibility of $X^*$, we get
\begin{align}
\label{eq.23}\|\mathcal{A}Z\|_2=\|\mathcal{A}X-\mathcal{A}X^*\|_2\leq\|\mathcal{A}X-b\|_2+
\|\mathcal{A}X^*-b\|_2\leq2\epsilon.
\end{align}

In the situation of $0<t<1$, by Lemma $\ref{le.9}$, we have
\begin{align}\label{eq.37}
\notag\left<\mathcal{A}Z^{(r)},\mathcal{A}Z\right>&\leq\|\mathcal{A}Z^{(r)}\|_2\|\mathcal{A}Z\|_2\\
\notag&\leq\sqrt{1+\delta_r}\|Z^{(r)}\|_F\|\mathcal{A}Z\|_2\\
\notag&=\sqrt{1+\delta_{\frac{1}{t}(tr)}}\|Z^{(r)}\|_F\|\mathcal{A}Z\|_2\\
\notag&\leq\sqrt{1+\left(\frac{2}{t}-1\right)\delta_{tr}}\|Z^{(r)}\|_F\|\mathcal{A}Z\|_2\\
&\leq\sqrt{\frac{1+\delta_{tr}}{t}}\|Z^{(r)}\|_F\|\mathcal{A}Z\|_2,
\end{align}
where in the first inequality, we used Cauchy-Schwarz inequality, and the second inequality follows from RIP of $r$-order.

Plugging (\ref{eq.23}) to (\ref{eq.37}), it follows that
\begin{align}\label{eq.38}
\left<\mathcal{A}Z^{(r)},\mathcal{A}Z\right>\leq 2\epsilon\sqrt{\frac{1+\delta_{tr}}{t}}\|Z^{(r)}\|_F.
\end{align}
Combining with equation (\ref{eq.16}) and inequalities (\ref{eq.19}), (\ref{eq.25}) and (\ref{eq.38}), we have
\begin{align}
\notag&\rho_{a,b}(t)t[t-(2-t)\delta_{tr}]\|Z^{(r)}\|^2_F\\
\notag&+4ab\epsilon t^2(2-t)\sqrt{\frac{1+\delta_{tr}}{t}}\|Z^{(r)}\|_F-t\bigg\{\{[(a+b)^2-4ab]t\\
\notag&-[(a+b)^2-2ab](2-t)\delta_{tr}\}\|Z^{(r)}\|^2_F\\
\notag&-2abr\delta_{tr}\alpha^2(2-t)\bigg\}\geq 0.
\end{align}
Applying inequality (\ref{eq.35}) to above equality, we get
\begin{align}\label{eq.39}
\notag&2abt(t-2)\bigg[(4-t)\left(\frac{t}{4-t}-\delta_{tr}\right)\|Z^{(r)}\|^2_F\\
\notag&-\left[2\epsilon\sqrt{(1+\delta_{tr})t}+\frac{4\delta_{tr}\|X-X^{(r)}\|_*}{\sqrt{r}}\right]\|Z^{(r)}\|_F\\
&-\frac{4\delta_{tr}\|X-X^{(r)}\|^2_*}{r}\bigg]\geq 0.
\end{align}
In the situation of $1\leq t<4/3$, due to the monotonicity of RIC $\delta_{tr}$, it implies that
\begin{align}\label{eq.40}
\notag\left<\mathcal{A}Z^{(r)},\mathcal{A}Z\right>&\leq\sqrt{1+\delta_r}\|Z^{(r)}\|_F\|\mathcal{A}Z\|_2\\
\notag&\leq\sqrt{1+\delta_{tr}}\|Z^{(r)}\|_F\|\mathcal{A}Z\|_2\\
&\leq 2\epsilon\sqrt{1+\delta_{tr}}\|Z^{(r)}\|_F.
\end{align}
It is easy to check that
\begin{align}\label{eq.41}
\notag ab&\geq \left(\frac{tr}{2}\right)^2-\frac{1}{4}=\frac{(2-t)^2r^2-1}{4}-(1-t)r^2\\
&>-(1-t)r^2.
\end{align}
Combining with equation (\ref{eq.17}) and inequalities (\ref{eq.19}), (\ref{eq.26}) and (\ref{eq.40}), it holds that
\begin{align}\label{eq.42}
\notag&\rho_{a,b}(t)\bigg\{\bigg[t(2-t)-(t^2-2t+2)\delta_{tr}\bigg]\|Z^{(r)}\|^2_F\\
\notag&-2r\alpha^2\delta_{tr}(t-1)\bigg\}+4\epsilon\sqrt{1+\delta_{tr}}t^3[ab-(t-1)r^2]\|Z^{(r)}\|_F\\
\notag&-(4-3t)\bigg\{\{[(a+b)^2-4ab]t\\
\notag&-[(a+b)^2-2ab](2-t)\delta_{tr}\}\|Z^{(r)}\|^2_F-2abr\delta_{tr}\alpha^2(2-t)\bigg\}\\
&\geq0.
\end{align}
Due to inequality (\ref{eq.35}), by fundamental calculation, we get
\begin{align}\label{eq.42}
\notag&2[(t-1)r^2-ab]t^2\bigg[(4-t)\left(\frac{t}{4-t}-\delta_{tr}\right)\|Z^{(r)}\|^2_F\\
\notag&-\left[2\epsilon\sqrt{1+\delta_{tr}}t+\frac{4\delta_{tr}\|X-X^{(r)}\|_*}{\sqrt{r}}\right]\|Z^{(r)}\|_F\\
&-\frac{4\delta_{tr}\|X-X^{(r)}\|^2_*}{r}\bigg]\geq 0.
\end{align}
Thereby, two second-order inequalities concerning $\|Z^{(r)}\|_F$ are established. Under the condition of $\delta_{tr}<t/(4-t)$, applying quadratic formula and some elementary compute, we have
\begin{align}\label{eq.43}
\notag\|Z^{(r)}\|_F\leq&\frac{1}{2(4-t)(\frac{t}{4-t}-\delta_{tr})}\bigg[\frac{4\delta_{tr}\|X-X^{(r)}\|_*}{\sqrt{r}}\\
\notag&+2\epsilon\sqrt{1+\delta_{tr}}(4-t)\kappa\\
\notag&+\bigg[\left(\frac{4\delta_{tr}\|X-X^{(r)}\|_*}{\sqrt{r}}
+2\epsilon\sqrt{1+\delta_{tr}}(4-t)\kappa\right)^2\\
\notag&+\frac{16\delta_{tr}\|X-X^{(r)}\|^2_*(4-t)}{r}\left(\frac{t}{4-t}-\delta_{tr}\right)\bigg]^{\frac{1}{2}}\bigg]\\
\notag\leq&\frac{1}{2(4-t)(\frac{t}{4-t}-\delta_{tr})}\bigg[\frac{8\delta_{tr}\|X-X^{(r)}\|_*}{\sqrt{r}}\\
\notag&+4\epsilon\sqrt{1+\delta_{tr}}(4-t)\kappa\\
\notag&+\frac{4\|X-X^{(r)}\|_*}{\sqrt{r}}\sqrt{(4-t)(\frac{t}{4-t}-\delta_{tr})\delta_{tr}}\bigg]\\
\notag=&\frac{2\sqrt{1+\delta_{tr}}\kappa}{\frac{t}{4-t}-\delta_{tr}}\epsilon\\
&+\frac{4\delta_{tr}
+2\sqrt{(4-t)(\frac{t}{4-t}-\delta_{tr})\delta_{tr}}}{(4-t)(\frac{t}{4-t}-\delta_{tr})\delta_{tr}}\frac{\|X-X^{(r)}\|_*}{\sqrt{r}}
\end{align}
with $\kappa$ is defined in Theorem $\ref{th.1}$, where the second inequality follows from the fact that for any vector $x\in\mathbb{R}^n$, $\|x\|_2\leq\|x\|_1$.

Then,
\begin{align}\label{eq.44}
\notag\|Z^{(r)}_c\|_F=&\left(\sum_{i\geq r+1}\sigma^2_i(U^{\top}ZV)\right)^{1/2}\\
\notag\leq&\left(\max_{i\geq r+1}\{\sigma_i(U^{\top}ZV)\}\sum_{i\geq r+1}\sigma_i(U^{\top}ZV)\right)^{1/2}\\
\notag=&\|Z^{(r)}_c\|^{1/2}\|Z^{(r)}_c\|^{1/2}_*\\
\notag\leq&\frac{\|Z^{(r)}\|^{1/2}_*}{\sqrt{r}}(\|Z^{(r)}\|_*+2\|X-X^{(r)}\|_*)^{1/2}\\
\leq&\left(\|Z^{(r)}\|^2_F+\frac{2\|X-X^{(r)}\|_*\|Z^{(r)}\|_F}{\sqrt{r}}\right)^{1/2},
\end{align}
where in the second inequality, we used Lemma $\ref{le.4}$, and the third inequality follows from the fact that for any $r$-rank matrix $X$, $\|X\|_*\leq \sqrt{r}\|X\|_F$.

A combination of (\ref{eq.43}) and (\ref{eq.44}) implies that
\begin{align}
\notag\|Z\|_F=&\left(\|Z^{(r)}_c\|^2_F+\|Z^{(r)}\|^2_F\right)^{1/2}\\
\notag\leq&\left(2\|Z^{(r)}\|^2_F+\frac{2\|X-X^{(r)}\|_*\|Z^{(r)}\|_F}{\sqrt{r}}\right)^{1/2}\\
\notag\leq&\sqrt{2}\|Z^{(r)}\|_F+\frac{\|X-X^{(r)}\|_*}{\sqrt{2r}}\\
\notag\leq&\frac{2\sqrt{2(1+\delta_{tr})}\kappa\epsilon}{\frac{t}{4-t}-\delta_{tr}}\\
\notag&+\frac{2\sqrt{2}}{\sqrt{r}}\left[\frac{1}{4}+\frac{2\delta_{tr}
+\sqrt{(4-t)(\frac{t}{4-t}-\delta_{tr})\delta_{tr}}}{\frac{t}{4-t}-\delta_{tr}}\right]\\
\notag&\times\|X-X^{(r)}\|_*.
\end{align}
In the situation of the error bound $\|\mathcal{A}^*(z)\|\leq\epsilon$, set $Z=X-X^{\circ}$. It holds that
\begin{align}
\notag\|\mathcal{A}^*\mathcal{A}Z\|&=\|\mathcal{A}^*(\mathcal{A}X-b)
-\mathcal{A}^*(\mathcal{A}X^{\circ}-b)\|\\
\notag&\leq\|\mathcal{A}^*(\mathcal{A}X-b)\|
+\|\mathcal{A}^*(\mathcal{A}X^{\circ}-b)\|\\
\notag&\leq2\epsilon.
\end{align}
Moreover,
\begin{align}
\notag\left<\mathcal{A}Z^{(r)},\mathcal{A}Z\right>&=\left<Z^{(r)},\mathcal{A}^*\mathcal{A}Z\right>\\
\notag&\leq\|Z^{(r)}\|_*\cdot2\epsilon\\
\notag&\leq2\epsilon\sqrt{r}\|Z^{(r)}\|_F.
\end{align}
The rest of steps are similar with the situation of the error bound $\|z\|_2\leq\epsilon$. The proof of Theorem $\ref{th.1}$ is completed.

\qed

\noindent \textbf{Proof of Theorem $\ref{th.2}$.} Let $E=\mbox{diag}(x)\in\mathbb{R}^{m\times m}$
with $$x=\frac{1}{\sqrt{2r}}(\underbrace{1,\cdots,1}_{2r},0,\cdots,0).$$
Define
$\mathcal{A}:~\mathbb{R}^{m\times m}\to\mathbb{R}^{q}$ as
$$\mathcal{A}X=\frac{2}{\sqrt{4-t}}\left(\sigma(X)-\left<\sigma(X),\sigma(E)\right>\sigma(E)\right).$$
Applying the Cauchy-Schwarz inequality, for all $\lceil tr\rceil$-rank matrices $X$, we get
\begin{align}
\notag\left|\left<\sigma(X),\sigma(E)\right>\right|&\leq\|\sigma(X)\|_2
\|\sigma(E)\cdot1_{\sup{(\sigma(X))}}\|_2\\
\notag&\leq\sqrt{\frac{\lceil tr\rceil}{2r}}\|X\|_F,
\end{align}
and
\begin{align}
\notag&\|\mathcal{A}X\|^2_2\\
\notag&=\frac{4}{4-t}\\
\notag&\times\left<\sigma(X)-\left<\sigma(X),\sigma(E)\right>\sigma(E),
\sigma(X)-\left<\sigma(X),\sigma(E)\right>\sigma(E)\right>\\
\notag&=\frac{4}{4-t}\left[\|X\|^2_F-|\left<\sigma(X),\sigma(E)\right>|^2\right].
\end{align}
Therefore,
\begin{align}
\notag\|\mathcal{A}X\|^2_2&\leq\left(1+\frac{t}{4-t}\right)\|X\|^2_F\\
\notag&\leq\left(1+\frac{t}{4-t}+\varepsilon\right)\|X\|^2_F.\\
\end{align}
For $r>1/\varepsilon$, we get
\begin{align}
\notag\|\mathcal{A}X\|^2_2&\geq\frac{4}{4-t}\left(1-\frac{\lceil tr\rceil}{2r}\right)\|X\|^2_F\\
\notag&\geq\frac{4}{4-t}\left(1-\frac{tr}{2r}-\frac{1}{2r}\right)\|X\|^2_F\\
\notag&\geq\frac{4}{4-t}\left(1-\frac{tr}{2r}-\frac{\varepsilon}{2}\right)\|X\|^2_F\\
\notag&\geq\left(1-\frac{t}{4-t}-\varepsilon\right)\|X\|^2_F.
\end{align}
Accordingly, by Definition $\ref{def.1}$, we obtain $\delta_{tr}=\delta_{\lceil tr\rceil}=\frac{t}{4-t}+\varepsilon$. Suppose
$Y_1=\mbox{diag}(y_1),~Y_2=\mbox{diag}(y_2)\in\mathbb{R}^{m\times m}$
with
$$y_1=(\underbrace{1,\cdots,1}_{r},0,\cdots,0)$$
and
$$y_2=(\underbrace{0,\cdots,0}_{r},\underbrace{-1,\cdots,-1}_{r},0,\cdots,0).$$
It is easy to verify that $Y_1$ and $Y_2$ are both matrices of rank $r$ such that $Y_1-Y_2\in\mathcal{N}(\mathcal{A})$, i.e., $\mathcal{A}Y_1=\mathcal{A}Y_2$. Consequently, it is not possible to reconstruct both $Y_1$ and $Y_2$ based on $(z,\mathcal{A})$.

\qed

\section{Conclusion}
\label{sec.6}

In this paper, we establish sufficient conditions which ensure the stable recovery or exactly recovery of any $r$-rank matrix satisfying a given linear system of equality constraints via solving a convex optimization problem, i.e., nuclear norm minimization. When the parameter $t$ is equal to $1$, the bound of RIC $\delta_r$ coincide with the result of \cite{Cai and Zhang}. Meanwhile, the derived upper bounds regarding the reconstruction error are better than those of \cite{Cai and Zhang}. Besides, the restricted isometry property condition is proved sharp. And integrated with the main results of \cite{Cai and Zhang 2014}, i.e., the case of $t>4/3$, for sharp RIP conditions
for all $t>0$, we present an intact characterization that can guarantee the exact recovery of all $r$-rank matrices by way of nuclear norm minimization. Furthermore, the numerical experiments demonstrate the performance of nuclear norm minimization method.


\ifCLASSOPTIONcaptionsoff
  \newpage
\fi



%

%

\begin{IEEEbiographynophoto}{JIANWEN HUANG}
was born in Gansu, China. He received the master's degree from the School of Mathematics and Statistics, Southwest University, Chongqing, China, in 2013. He is currently
pursuing the Ph.D. degree with the School of Mathematics and Statistics, Southwest University.
His research interests include machine learning, data mining, sparse learning and extreme value theory.
\end{IEEEbiographynophoto}

\begin{IEEEbiographynophoto}{JIANJUN WANG}
received the BS degree in mathematical education from NingXia University in 2000 and the MS degree in fundamental mathematics in 2003 from NingXia University, China. And Ph.D degree in applied mathematics was obtained from the Institute for Information and System Science, Xi¡¯an Jiaotong University in Dec. 2006. He is currently a professor in the School of Mathematics \& Statistics at at Southwest University of China. His research focus on machine learning, data mining, neural networks and sparse learning.
\end{IEEEbiographynophoto}

\begin{IEEEbiographynophoto}{FENG ZHANG}
 was born in Sichuan, China. He received the master's degree from the School of Mathematics and Statistics, Southwest University, Chongqing, China, in 2017. He is currently
pursuing the Ph.D. degree with the School of Mathematics and Statistics, Southwest University.
His research focus on compressed sensing (CS), low-rank tensor approximation (LRTA).
\end{IEEEbiographynophoto}

\begin{IEEEbiographynophoto}{WENDONG WANG}
received the B.S. degree in mathematics and applied mathematics from Weinan Normal University, Weinan, China, in 2011, the M.S. degree in applied mathematics and the Ph.D. degree in statistics from Southwest University, Chongqing, China, in 2014 and 2017, respectively. His research interests include compressed sensing, low-rank matrix recovery and machine learning. He is currently doing postdoctoral research at Southwest University, Chongqing, China.
\end{IEEEbiographynophoto}




\end{document}